\documentclass{emulateapj}

\usepackage{natbib}
\usepackage{longtable}
\newcommand{\chandra}{{\emph{Chandra}}}
\newcommand{\rxj}{RX~J1131$-$1231}
\newcommand{\rxjs}{J1131}

\newcommand{\flux}{\hbox{ergs~cm$^{-2}$~s$^{-1}$}}
\newcommand{\lumin}{\hbox{ergs~s$^{-1}$}}

\newcommand{\delayABonesig}{$11.98^{+1.52}_{-1.27}$}
\newcommand{\delayAConesig}{$ 9.61^{+1.97}_{-1.57}$}
\newcommand{\delayADonesig}{$-87 \pm 8$}
\newcommand{\delayBConesig}{$-2.20^{+1.55}_{-1.64}$}

\shorttitle{Time-Delay Measurement for the Quadruple Lens RX J1131$-$1231}
\shortauthors{Morgan et al.}

\begin{document}

\title{Time-Delay Measurement for the Quadruple Lens RX J1131$-$1231
\altaffilmark{1}}

\author{N.~D.~Morgan\altaffilmark{2},
	C.~S.~Kochanek\altaffilmark{2},
	E.~E.~Falco\altaffilmark{3},
	X.~Dai\altaffilmark{2}
}

\altaffiltext{1}{Based on observations obtained with the Small and Moderate 
Aperture Research Telescope System (SMARTS) 1.3m operated by the SMARTS 
Consortium, the NASA/ESA {\it Hubble Space Telescope}, and the Spitzer Space 
Telescope.  {\it HST} observations are obtained at the Space Telescope Science
Institute, which is operated by the Association of Universities for Research 
in Astronomy, Inc., under NASA contract NAS 5-26555.  These observations are 
associated with {\it HST} program 9744.  The Spitzer telescope is operated by 
the Jet Propulsion Laboratory, California Institute of Technology, under a 
contract with NASA.  Support for this work was provided by NASA through an 
award issued by JPL/Caltech.  These observations are associated with Spitzer 
program 20451.}

\altaffiltext{2}{Department of Astronomy, Ohio State University, Columbus, OH 
43210; nmorgan@astronomy.ohio-state.edu, ckochanek@astronomy.ohio-state.edu,
xinyu@astronomy.ohio-state.edu}

\altaffiltext{3}{Harvard-Smithsonian Center for Astrophysics, 60 Garden 
Street, Cambridge, MA 02138; falco@cfa.harvard.edu}

\begin{abstract}

We have measured the three time delays for the quadruple gravitational lens 
\rxj\ using two seasons of monitoring data.  The short delays between cusp 
images are A--B = \delayABonesig\ days and A--C = \delayAConesig\ days.  The 
long A--D delay for the counter image is not as precisely determined because 
of the season gaps, but the data suggest a delay of \delayADonesig\ days.  The
short delays are difficult to explain using standard isothermal halo models 
of the lensing potential, which instead prefer A--B and A--C delays of
$\sim 1$ day for reasonable values of the Hubble constant.  Matching the
cusp delays is possible by adding a significant ($\sim 5 \times 
10^{10}$ M$_{\sun}$) amount of matter nearly coincident ($\sim 0\farcs05$ 
South-East) with the A image.  Adding such a satellite also helps
improve the quasar and lens astrometry of the model, reduces the velocity 
dispersion of the main lens and shifts it closer to the Fundamental Plane.  
This is suggestive of a satellite galaxy to the primary lens, but its 
expected luminosity and proximity to both image A and the system's bright 
Einstein ring make visual identification impossible even with the existing 
{\it HST} data.  We also find evidence for significant 
structure along the line of sight toward the lens.  Archival {\it Chandra} 
observations show two nearby regions of extended X-ray emission, each with 
bolometric X-ray luminosities of 2-3 $\times$ $10^{43}$ ergs s$^{-1}$.  The 
brighter region is located $\Delta \theta \approx$ 153\arcsec\ from the lens 
and centered on a $z=0.1$ foreground cD galaxy, and the fainter and 
presumably more distant region is 4-5 times closer (in angular separation) 
to the lens and likely corresponds to the weaker of two galaxy red sequences 
(which includes the lens galaxy) previously detected at optical wavelengths.

\end{abstract}


\section{Introduction}

Gravitationally lensed quasars are useful tools for studying the matter 
distribution of intermediate redshift ($0.1 \lesssim z \lesssim 1.0$) 
galaxies.  Robust estimates of the total (luminous plus dark) lensing 
mass usually follows from just measuring the image positions.  However,
quantifying the 
presence of halo substructure, constraining the mass profile's radial 
distribution, or parsing the relative contributions from stars and dark matter
requires additional information.  For example, the flux ratios of lensed
images are sensitive to the presence of halo substructure (e.g., 
\citealt{2002ApJ...567L...5M}) and the incidence of so-called ``anomalous'' 
ratios at radio wavelengths is consistent with CDM predictions that a few 
percent of the halo mass should be left in satellites 
\citep{2002ApJ...572...25D}.  At optical wavelengths, interpreting
such anomalies in light of stellar microlensing implies that $\lesssim 25\%$ 
of the surface mass density at the image locations must be in a smooth 
(non-stellar) form \citep{2002ApJ...580..685S, 2006ApJ...639....1K}.

Another key observation is time delays.  The time delays between lensed
images measure a combination of the Hubble constant
and the surface mass density $\langle \kappa \rangle = \langle \Sigma \rangle
 / \Sigma_c$ of the lens at the radius of the lensed images, with 
$H_{\rm o} \propto (1 - \langle \kappa \rangle)$ to lowest order
\citep{2002ApJ...578...25K}.  For the six simple time delay lenses that we 
have at present, \citet{2002ApJ...578...25K} noted that four (PG1115+080, 
\citealt{1997ApJ...475L..85S}, \citealt{1997ApJ...489...21B}, 
\citealt{1998ApJ...509..551I}; SBS1520+530, \citealt{2002A&A...391..481B}; 
B1600+434, \citealt{2000ApJ...544..117B}, \citealt{2000A&A...356..391K}; and 
HE2149--2745, \citealt{2002A&A...383...71B})
yield $H_{\rm o}$ predictions consistent with the {\it HST} Key Project 
result of 72 km s$^{-1}$ Mpc$^{-1}$ \citep{2001ApJ...553...47F}
only if the lensing galaxies have constant $M/L$ profiles (that is, declining 
rotation curves at the system's Einstein radius).  
The remaining two systems (HE1104--1805, \citealt{2003ApJ...594..101O}; 
HE0435--1223, \citealt{2006ApJ...640...47K}) have measured delays consistent
with the Key Project only with mass profiles that drop 
off {\it slower} than isothermal (that is, {\it rising} rotation curves at the 
locations of the lensed images).  In the case of HE0435--1223, the best-fit 
model yields an average mass convergence 
at the system's Einstein radius that is 20\% larger ($\langle\kappa\rangle 
\simeq 0.6 \pm 0.05$) than expected for an isothermal halo.
Taken at face value, the range in $\langle\kappa\rangle$ estimates imply
a heterogeneous mix of mass profiles for lensing galaxies, with the likely
explanation that we are seeing the effects of the lensing environment.  
Lens galaxies located at the center of their respective groups will naturally 
have an extended dark matter halo (yielding flat or rising rotation curves), 
while lens galaxies located in the field or group periphery will have mass 
distributions dominated by their stellar components (falling rotation curves).
Under this interpretation, it becomes important to not only expand the sample 
of systems with measured time-delays in order to sample as wide a range of
environments as possible, but also to characterize the lensing environment 
for systems with measured delays as thoroughly as possible.

In this paper, we present our time-delay measurements and explore the
lensing environment for the quadruple lens \rxj\ (hereafter \rxjs).  Our 
original motivation was to use the measured delays and assumed $H_{\rm o}$
to estimate the lens surface mass density at the \rxjs\ image positions.  
Coupled with the robust total mass estimate obtained from the image 
separations and the measured luminosity distribution of the lens obtained from 
{\it HST}/NICMOS data, this would allow us to estimate the radial mass profile of the
lens and separate the luminous and dark contributions to the total lensing 
mass (as done recently for HE~0435--1223; \citealt{2006ApJ...640...47K}).  
Unfortunately, the measurement of unexpected delays for this system makes 
such an approach unfeasible.  Instead, we simply focus on constructing a 
plausible macromodel that can reproduce the delays while assuming an 
isothermal halo for the main lens.  

This system was originally identified as a {\it ROSAT} X-ray source and 
subsequently revealed to be a quadruply-imaged quasar by 
\citet{2003A&A...406L..43S}.  The lensing geometry consists of a background 
$z_s=0.658$ quasar lensed by a foreground $z_l=0.295$ elliptical galaxy, and 
there is a prominent Einstein ring evident even in ground-based optical data.  
\citet{2003A&A...406L..43S} also detected variability on the order of 0.3 
magnitudes from the total integrated system flux over an eight month baseline.
Overall, the system's demonstrated variability, wide image separation (A to D 
separation of 3\farcs2), prominent lens galaxy and quad plus ring morphology 
made \rxjs\ an attractive target for long-term monitoring and followup 
studies.  In \S\ref{sec:monitoring} and \S\ref{sec:delays} we describe our 
monitoring program and time-delay measurements and comment on the microlensing
variability observed in the system over the course of the campaign.  
High-resolution {\it HST}/ACS and NICMOS images of the lens and immediate 
environment are described in \S\ref{sec:hst}, and evidence for at least two 
clusters along the line of sight are presented using far IR 
(\S\ref{sec:spitzer}) and X-ray (\S\ref{sec:chandra}) observations obtained 
with the {\it Spitzer} and {\it Chandra} Space Telescopes.  We then explore 
several iterations of parametric and non-parametric lens models and discuss 
their implications in \S\ref{sec:models}.  In summary, we find it difficult 
to simultaneously explain the system geometry and measured time delays 
using a single profile for the main lens galaxy, but find it possible if a 
significant amount of substructure is present near one of the cusp images.  
Finally, we summarize our results and describe avenues for future work on this
system in \S\ref{sec:conclusions}.

\section{Lens Monitoring}\label{sec:monitoring}

The monitoring data for J1131 were obtained as part of the ongoing program 
described by \citet{2006ApJ...640...47K};  details of the reduction pipeline 
and lens fitting procedures can also be found in that reference.  Briefly, we 
monitor approximately 25 lensed quasars at cadences of 1-2 times per week, 
with typical observations of three 2-5 minute exposures per epoch in either 
Sloan $r$ or Johnson $R$.  Each lens is assigned up to 5 nearby reference 
stars which serve as local flux calibrators and as shape templates for the 
analytic PSF model.  Point sources are modeled as a combination of three 
elliptical Gaussians and extended objects (lens galaxies) are modeled using 
Gaussian approximations to de Vaucouleurs profiles (which allows for rapid 
analytic convolutions with the seeing disc).  The relative astrometry for each 
system is fixed using observations from the CASTLES\footnote{Due to the career
moves of the investigators, CASTLES is no longer an acronym but rather just 
the name of the survey.} {\it HST} program, and relative photometry for the 
system images are obtained by minimizing $\chi^2$ residuals between the data 
and analytic model.

The monitoring data for \rxjs\ were obtained using the queue-scheduled 
SMARTS/CTIO 1.3m telescope in the Southern Hemisphere and the MDM 2.4m, 
WIYN 3.5m, and APO 3.5m telescopes in the Northern Hemisphere.  The bulk 
(94\%) of data were taken with the SMARTS 1.3m using the dual-beam optical/IR 
ANDICAM camera \citep{2003SPIE.4841..827D} and covering the 2003-2004 and 
2004-2005 seasons for a total of 101 epochs.  These observations consisted of 
three 5 minute $R$-band exposures and six 2.5 minute $J$-band exposures at 
each epoch.  The IR data are considerably less sensitive than the optical data 
and we focus only on the $R$-band light curves here.  Additional observations 
include three Sloan $r$-band epochs consisting of 1.5-2 minute exposures using
the SPIcam CCD on the APO 3.5m, two $R$-band epochs of 2-3 minute exposures 
using the WTTM camera on the WIYN 3.5m, and two $R$-band epochs of 3 minute 
exposures using the 8K Mosaic CCD on the MDM 2.4m.  The APO, WIYN and MDM 
light curves were calibrated by interpolating the quasar and reference star 
data points onto the respective SMARTS curves, yielding star offsets of 
$-0.01$, $0.00$ and $-0.02$ mags and quasar offsets of $0.08$, $-0.02$ and 
$0.04$ mags, respectively.  We only kept data with seeing better than 1\farcs7
FWHM, or about half the maximum image separation.

\begin{figure*}[t]
\plotone{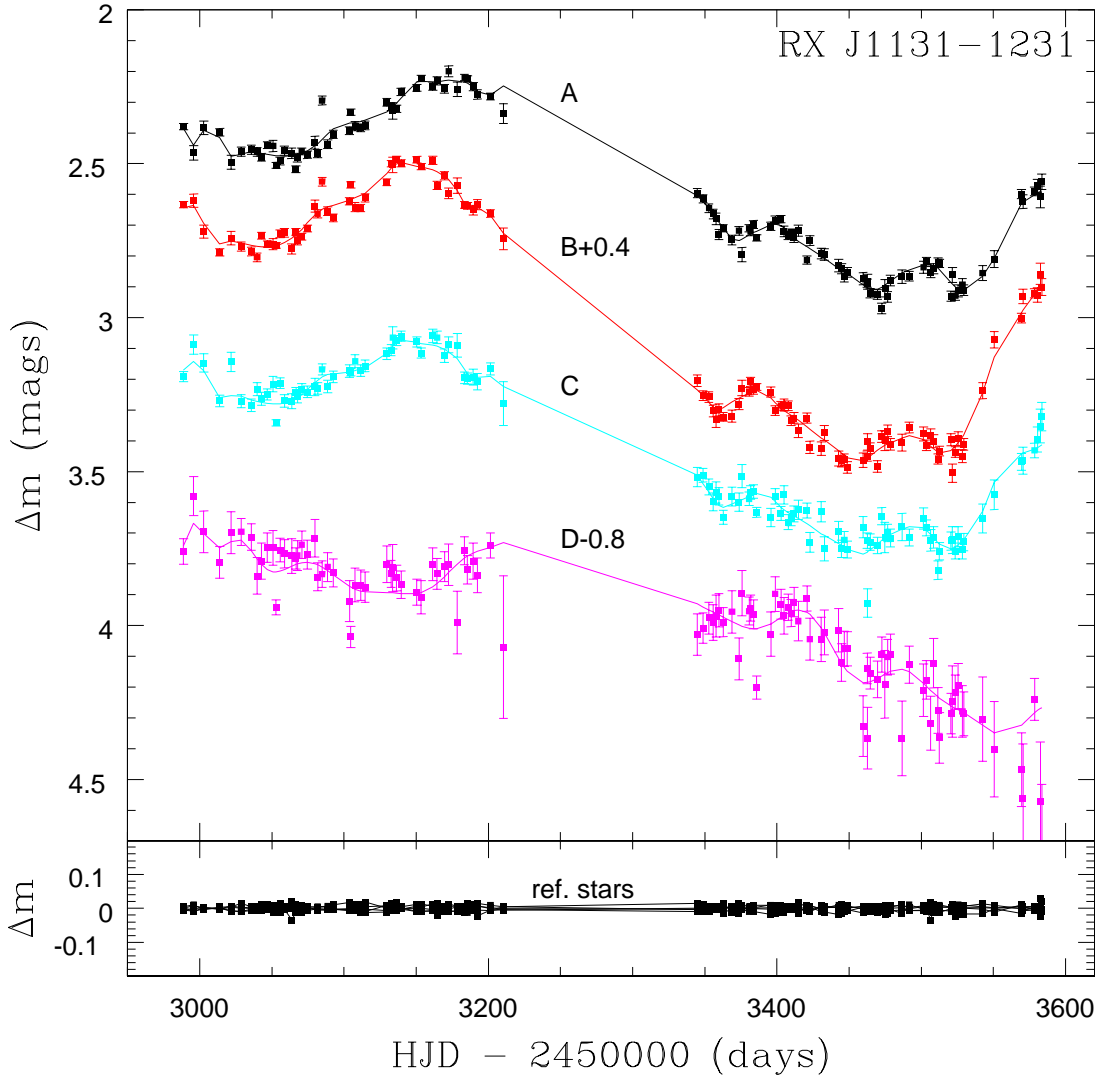}
\caption{Light curves for images A--D.  The B and D curves are offset for
clarity.  The best fit models are for $N_{\rm src}=30$, $N_{\mu}=4$ for each 
separate season for images $A$, $B$ and $C$, and for $N_{\rm src}=40$, 
$N_{\mu}=3$ for the joint seasons for image D.  Differential photometry for
5 nearby reference stars is shown in the bottom panel.}  
\label{fig:curves}
\end{figure*}

Figure~\ref{fig:curves} shows the four quasar light curves and 
Table~\ref{tab:lcurve} gives the quasar magnitudes and reference star 
differential variability at each epoch.  Over the two seasons, each image 
varied by as much as 0.7 mag.  Significant correlated variability on 
timescales of weeks is also apparent for the three brighter images.

\section{Time Delays Measurements}\label{sec:delays}

The light curves are modeled as a sum of two Legendre polynomial series.  The 
first series, of order $N_{\rm{src}}$, models the intrinsic source variability
of the quasar and is the same for all four quasars images up to a time delay 
shift for each pair.  The second series, of order $N_{\mu}$, is different for 
each image and models the different macrolensing magnifications ($0^{th}$ 
order) and the uncorrelated microlensing variability for each quasar image 
($1^{st}$ order and above).  Exact prescriptions for the fitting forms can
be found in \citet{2006ApJ...640...47K}.  The merit function for each light curve is minimized
by differentiating with respect to the polynomial coefficients and solving
the resulting linear set of equations, which yields estimates for the
time delays between each image, a lightcurve for the intrinsic source 
variability, and the relative microlensing lightcurves for three of the four
images.

\begin{figure}
\plotone{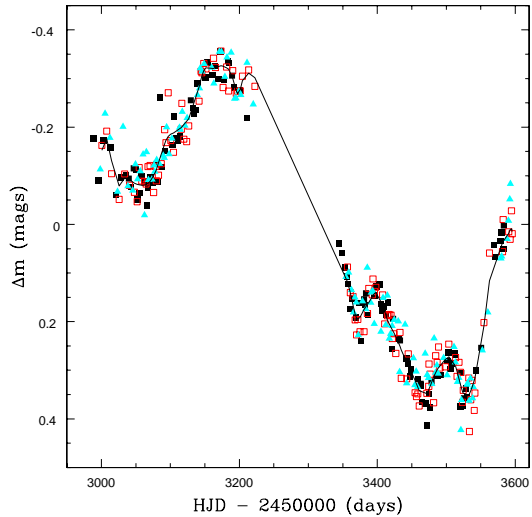}
\caption{Overlapping curves for images A (open squares), B (triangles) and 
C (filled squares) after shifting by the best-fit time delay and removing 
the macromodel magnification and microlensing variability from images B 
and C.  The much noisier D curve is not shown.  Errorbars are suppressed 
for clarity.}  
\label{fig:shifted}
\end{figure}

The A, B and C images show relative delays of about 1-2 weeks, much 
shorter than the length of each observing season.  Since all the 
signal when solving for the ABC delays will come from same-season 
data, we fit the two season's ABC light curves simultaneously (that is,
we solve for a single set of time delays constrained by both seasons) but 
model each season's lightcurve separately.  (We discuss modeling the D curve
below).  The $F$-test is used to decide when further increases 
in $N_{\rm{src}}$ and $N_{\mu}$ do not yield improvements to the fits.  We 
find that most of the source-curve improvements occur up to $N_{\rm{src}} 
= 10$, beyond which the higher source-curve complexity does not significantly 
improve $\chi^2$.  Significant improvement in the microlensing model occurs 
from $N_{\mu}=0$ (macromodel magnifications only and no microlensing) to 
$N_{\mu}=4$, arguing for microlensing variations that are somewhat more 
complex than just a simple linear trend (which would be described by 
$N_{\mu}=1$).  We adopt $N_{\rm{src}} = 30$, $N_{\mu}=4$ as our standard model 
for the A, B and C images, where the larger source curve order should
lead to conservative uncertainties in the delay estimates.  This model gives 
A--B, A--C and B--C delays of \delayABonesig\, \delayAConesig\ and 
\delayBConesig\ days (1$\sigma$), respectively.  As a 
check on the results, we also fitted each season's light curve as a 
stand-alone dataset using a separate code that uses brute-force (Powell) 
minimization, and we find delays consistent with the quoted error bars.  
Figure~\ref{fig:shifted} shows the ABC light curves after shifting by the 
relative delays and subtracting the microlensing and macromodel 
magnifications, leaving just the source variability observed in each image.

Fitting for image D's light curve is less precise.  In addition to the lower 
signal-to-noise, D is expected to lag the other images by $\sim100$ days 
\citep{2003A&A...406L..43S}, which would shift about half of D's data points 
into the season gaps of the other images.  This leaves less overlapping data 
to measure a delay, but we can improve the available signal by modeling the 
two seasons as one continuous light curve.  Fixing the relative ABC delays at 
the values determined above, we use a $N_{\rm{src}} = 40$, $N_{\mu}=3$ model 
to obtain an A--D delay of \delayADonesig\ days ($1\sigma$).  The reported 
best-fit is driven by aligning the broad valley in image D at MJD $\approx$ 
3125 days with the similar valley in the ABC curves at MJD $\approx$ 3040, 
with most of the season two drop in D's brightness shifted into the ABC 
season gap.  If this feature is not correctly matched, then the absence of 
other distinguishing features in the D curve would mean that the A--D delay is 
unconstrained with the extant data.  If the feature is correctly matched, 
then the D delay is measured to an accuracy of about 10\%.

\begin{figure}[b]
\plotone{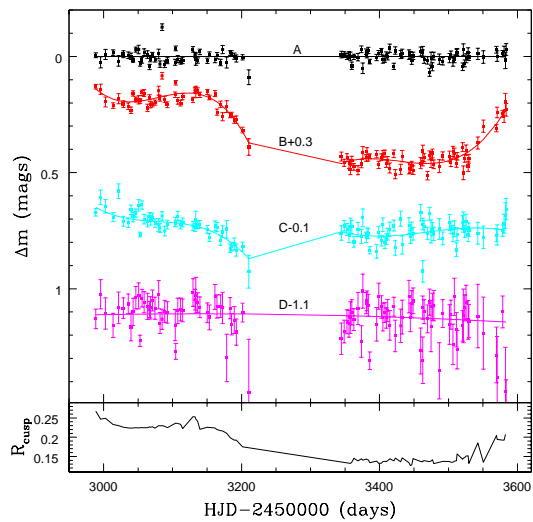}
\caption{({\it Top}):  The microlensing light curves for \rxj.
Shown are the light curves for images A--D after subtracting the time-delay 
shifted A model curve as a model for the source variability.  ({\it Bottom}):
$R_{\rm cusp}$ for images A, B and C formed using the best-fit curves in 
Figure \ref{fig:curves} after shifting by the respective time delays.}  
\label{fig:micro}
\end{figure}

Figure \ref{fig:micro} shows the residual microlensing light curves after 
subtracting the appropriately time-delay shifted source variability 
(arbitrarily set to be image A's light curve) from each quasar image.  Both B 
and C show a drop in brightness relative to A toward the end of the first 
season, and image B shows a rapid rise in brightness at the end of the second 
season.  Image D does not show significant variability apart from a gradual 
($\Delta m \simeq 0.05$ mag) decline during both seasons, although the lower 
signal to noise complicates the analysis.  

It is likely that image A is responsible for the first season drop seen in the
B and C curves since we only measure the differential effects.  This would 
imply a net increase in A's brightness by $\sim$0.2 mag due to microlensing 
toward the end of the first season.  The differences between the B and C 
microlensing light curves are evidence for a similar amount of microlensing in
one or both of these images as well.  For example, B's upswing at the end of 
season two is not seen in image C.  Also, image B spends most of season two 
$\sim$0.2 mag fainter than its season one average, while image C maintains 
more or less the same average flux from one season to the next.

One succinct way to quantify the effects of microlensing is to look at the 
lensing cusp relation \citep{1998MNRAS.295..587M}, which states that the sum 
of the signed image magnifications is zero for the triplet of images formed 
when a source is near a cusp caustic (images A, B and C for \rxjs).  The 
relation can be phrased in terms of the observed image fluxes ($F_A, F_B, 
F_C$) by dividing through by the unknown source flux, yielding $R_{\rm cusp} 
= |F_A + F_B + F_C| / (|F_A| + |F_B| + |F_C|)$.  While the cusp relation
for a source asymptotically close to a cusp has $R_{\rm cusp} = 0$, in 
practice $R_{\rm cusp}$ as defined will increase by several tenths as the 
source is placed farther from the cusp, not to mention the additional 
perturbing effects of micro- and millilensing 
\citep[hereafter KGP]{2003ApJ...598..138K}.  For a system configuration such 
as \rxjs, one can place a strong upper limit on the baseline $R_{\rm cusp}$ 
value (before micro- or millilensing perturbations, that is, assuming a 
perfectly smooth potential model) of 
$\lesssim$0.1 (KGP).  This value is significantly lower than the 
$R_{\rm cusp} \approx 0.35 \pm 0.03$ measured from the 
\citet{2003A&A...406L..43S} $V$-band discovery data, and lead KGP to classify 
\rxjs\ as a ``cusp-anomaly'' system.

\begin{figure*}[t]
\plotone{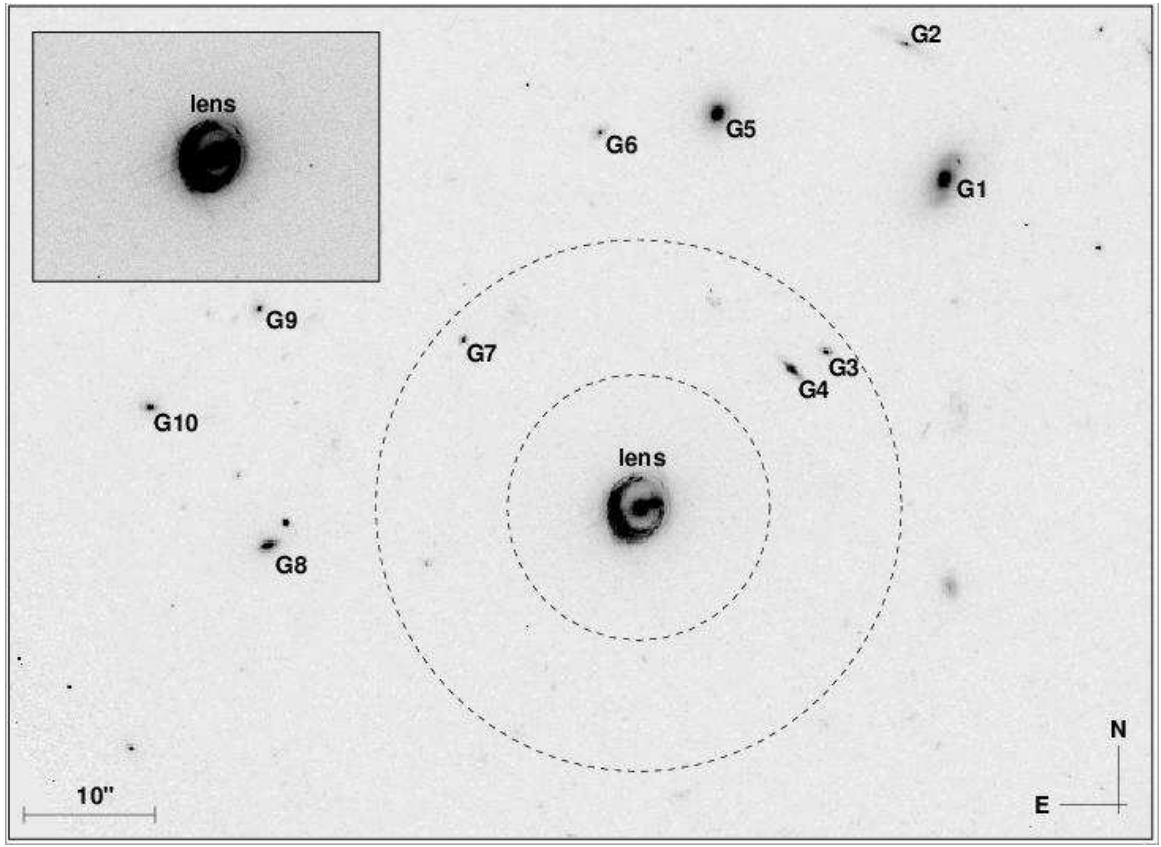}
\caption{PyDrizzled {\it HST}/ACS $I$-band image of \rxj\ and the surrounding 
field.  Prominent nearby galaxies are labeled.  Concentric circles mark the 
10\arcsec\ and 20\arcsec\ radii centered on the lens galaxy.  The inset 
shows the inner 10\arcsec\ region at the same scale but at 10$\times$ higher 
contrast.}\label{fig:hstwide}
\end{figure*}

Anomalous $R_{\rm cusp}$ values indicate the presence of substructure in the
lens galaxy, but it can not discern whether the anomaly is due to microlensing 
by stars or due to millilensing by substructure (either dark matter clumps or 
luminous satellites).  However, only microlensing can change $R_{\rm cusp}$ 
on observable timescales, so tracking changes in $R_{\rm cusp}$ tracks the 
microlensing contributions.  The bottom panel of Figure 
\ref{fig:micro} shows $R_{\rm cusp}$ computed using the time-delay corrected 
light curves (shifting to the image A time frame).  $R_{\rm cusp}$ 
is $\sim$0.23 during most of the first season but begins to drop, becoming
less ``anomalous'' with respect to the KGP estimate toward the season gap.
At the same time, B and C drop in brightness and the $R_{\rm cusp}$ value
remains at $\sim$0.14 for the bulk of the second season.  The 
resolved {\it Chandra} X-ray observations of J1131 reported by 
\citet{astro-ph/0509027} have $R_{\rm cusp} = 0.75 \pm 0.05$ on MJD = 3108 due
to the significantly dimmed flux of component A (see their Figure~1a).  
Taken together, the SMARTS and {\it Chandra} data suggest that the saddlepoint
A image was in a demagnified microlensing state for much of season one and 
varied closer to the unperturbed macromodel magnification heading into season 
two.  

The {\it HST} observations of \rxjs\ described in the next section yield 
respective $R_{\rm cusp}$ values in the V, I and H filters of $0.26 \pm 0.06$, 
$0.17 \pm 0.05$ and $0.08 \pm 0.01$.  Assuming the baseline $R_{\rm cusp}$ value
of $\lesssim 0.1$ (KGP), the trend with wavelength agrees with 
the notion that microlensing preferentially affects smaller emission regions 
of the quasar accretion disk (under the assumption that longer wavelengths 
correspond to lower disk temperatures, which then correspond to larger radii 
from the central black hole).  However, it is important to remember that the 
cusp anomaly may be due to the presence of substructure on scales 
comparable to the image separations.  For example, models with a satellite 
galaxy near one of the cusp images (explored in \S\ref{sec:substruc}) predict 
an $R_{\rm cusp}$ value of about 0.5 even before any microlensing effects.  
In such a case, the offsets from $R_{\rm cusp} \lesssim 0.1$ 
cannot be interpreted as strictly a sign of microlensing, but the
variations in $R_{\rm cusp}$ seen in time and across wavelength still point 
to abundant microlensing for \rxjs.
\ \\

\section{HST Observations}\label{sec:hst}

\begin{figure*}[t]
\plotone{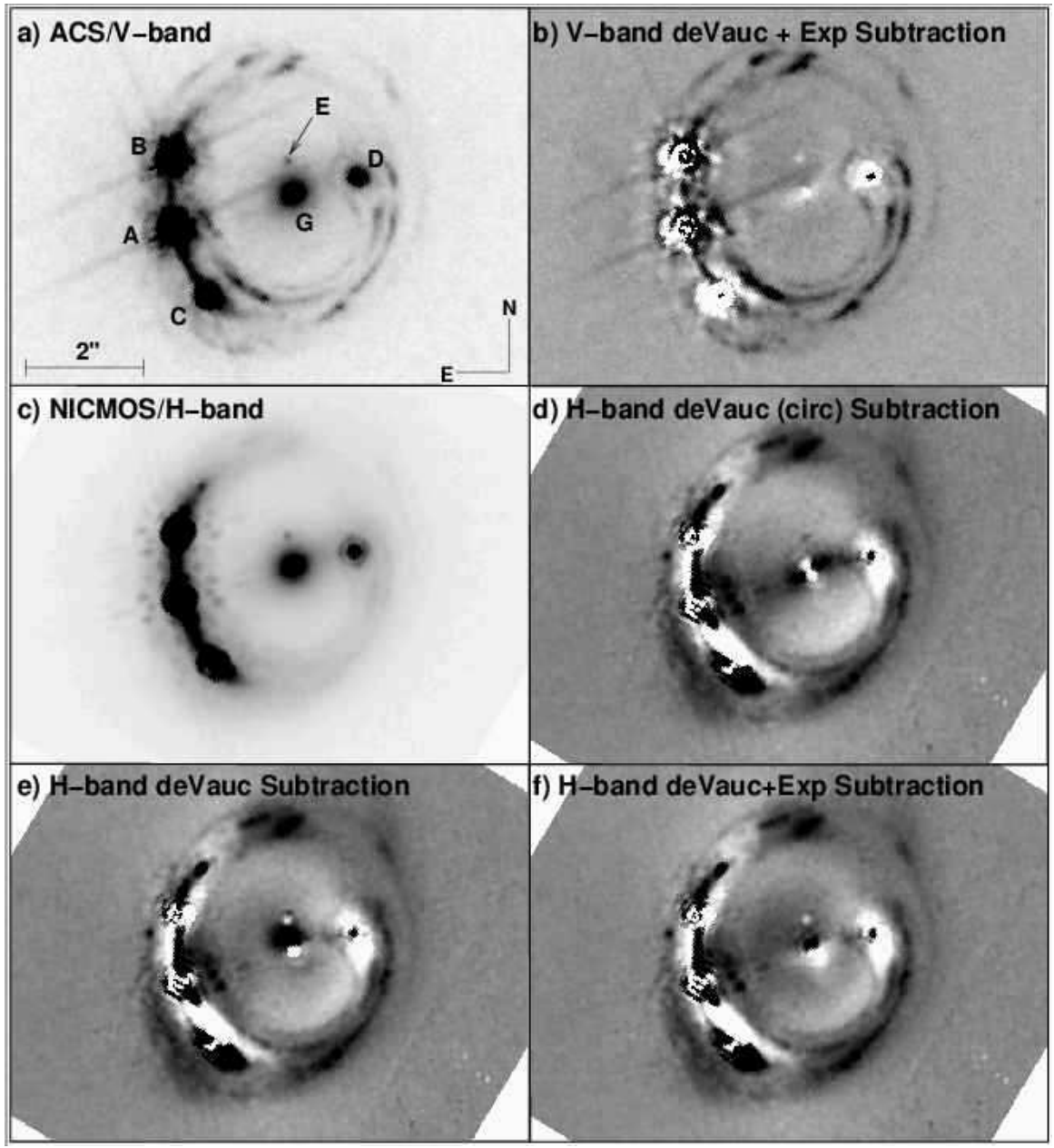}
\caption{Combined {\it HST} images of \rxj. {\it (a)}: ACS/$V$-band image.  
{\it (b)}: Same as {\it (a)}, but after subtracting the best-fit point source,
lens galaxy and ring model.  {\it (c)}: NICMOS/$H$-band image.  {\it (d)}: 
Same as {\it (c)}, but after subtracting the best-fit point source, lens 
galaxy and ring model.  Contrast in both residual panels stretches from $-7\%$ 
to $+7\%$ of the respective peak lens galaxy flux.}
\label{fig:hstclose}
\end{figure*}

High-resolution images of \rxjs\ were obtained with the ACS/Wide Field Camera 
(WFC; \citealt{1998SPIE.3356..234F}) and NICMOS detectors on board the {\it 
HST} as part of the CASTLES imaging program (PID 9744).  Both of these public
data sets have also been recently analyzed by \citet{astro-ph/0602309}, and 
we make comparisons between the separate reductions below.  The optical 
ACS images were taken with the F814W ($I$-band) and F555W ($V$-band) filters 
on 22 and 24 June 2004, respectively.  Exposures were 33 minutes through each 
filter using a 5-point dither pattern.  The data were reduced using the 
IRAF/CALACS package as part of the ``on-the-fly'' reprocessing at the time 
of download.  Subsequent cosmic-ray rejection, geometric correction, and 
image combinations were performed using the standard PyRAF programs available 
for ACS data reduction.  The infrared NICMOS images were taken with the NIC2 
camera and F160W ($H$-band) filter on 17 November 2003, and consisted of two 
sets of four-point dither MULTIACCUM exposures for a total on-source 
integration of about 89 minutes.  The NICMOS data were reduced using the 
NICRED software described in \citet{2000ApJ...536..584L}.

Figure~\ref{fig:hstwide} shows a 1\arcmin\ field of view surrounding \rxjs\ 
from the stacked $I$-band data.  The environment within 10\arcsec\ of the 
lens is devoid of galaxies, and the nearest galaxies to the lens (G3, G4 
and G7) are 15-20\arcsec\ distant.  The Figure~\ref{fig:hstwide} inset shows 
the inner 10\arcsec\ region centered on \rxjs\ at 10$\times$ higher contrast 
to emphasize the lack of nearby structure.  The nearest galaxies with 
comparable magnitudes to the lens are G1 and G5, roughly 30\arcsec\ distant 
toward the North-North-West.

In Figures~\ref{fig:hstclose}a and \ref{fig:hstclose}c we show the $V$- and 
$H$-band closeup images of \rxjs.  A complex pattern of lensed knots from the 
quasar host galaxy is present, presumably because the relatively low 
redshift host galaxy is a star-forming spiral galaxy.  There is also a faint 
fifth object, component E in Figure~\ref{fig:hstclose} (denoted component X 
in \citealt{astro-ph/0602309}), roughly 0\farcs5 North of the lens galaxy.
The object shows a clear Airy ring in the $H$-band data, so it is unresolved.

To model the light distribution, we first tried fitting the $H$-band data 
using a seven component model: five point-sources for images A-E, a de 
Vaucouleurs profile for the lens galaxy, and a lensed de Vaucouleurs 
profile for the quasar host galaxy to model the continuous portion of the 
Einstein ring emission.  For the fitting, we used the {\tt IMFIT} package 
written by B. McLeod (see, e.g., \citealt{2000ApJ...536..584L}).  The lensing 
potential was modeled as a singular isothermal sphere plus external shear 
using just the $H$-band image positions as constraints for the purpose of 
mapping the host galaxy onto the image plane.  We also tried exponential and 
Gaussian profiles for the quasar host, but the de Vaucouleurs profile yielded 
the lowest residuals.  Figure \ref{fig:hstclose}d shows the residuals after 
fitting a circularly symmetric de Vaucouleurs profile for the lens galaxy and 
reveals the lens to be elongated roughly along the A-D line.  Allowing an 
elliptical de Vaucouleurs profile for the lens does well at modeling the 
galaxy ellipticity (Figure \ref{fig:hstclose}e) but still leaves extended 
residuals near the center of the lens with a peak flux of about 12\% of the 
unsubtracted galaxy peak flux.  To model this extra emission, we added an 
exponential profile centered on the de Vaucouleurs model.  This addition did 
a better job at accounting for the bulk of the extra emission, although there 
are systematic residuals in the core of the lens galaxy ($\sim$10\% of the 
unsubtracted galaxy flux; Figure \ref{fig:hstclose}f) that are suggestive of 
a slightly misaligned center for the de Vaucouleurs and exponential profiles.  

The inability to model the main lens with a single profile was noted by 
\citet{astro-ph/0602309} as well.  The two-component galaxy model naturally 
suggests a disk+bulge morphology, except that we find the de Vaucouleurs 
model to be a better fit to the elongated structure than to the bulge.  For an
S0 galaxy, one would expect the opposite, namely an exponential disk and a de 
Vaucouleurs bulge.  We tried to coax the models to assign the exponential 
profile to the disk by using judicious choices for the initial conditions, 
but the models consistently preferred a de Vaucouleurs disk and exponential 
bulge.  The de Vaucouleurs component of the galaxy model is best characterized
by an effective radius $R_e = 1\farcs58 \pm 0\farcs29$, axis ratio of 
$0.51 \pm 0.02$, and major axis orientation of $-62\degr \pm 2\degr$.  The 
exponential component of the galaxy has a best-fit scale length of 
0\farcs20 $\pm$ 0.01, axis ratio of 0.79 $\pm$ 0.04, and major axis 
orientation 33\degr\ $\pm$ 4\degr.  

For the $V$- and $I$-band data, we fixed the astrometric and structural 
components of the system at the values from the $H$-band eight component 
model and solved for the relative photometry.  Table~\ref{tab:astromphot} 
lists the results for all three filters.  The relative positions for the 
A, B and C images agree within the quoted errorbars with the 
\citet{astro-ph/0602309} positions, but we do find a small offset of 
0\farcs009 (3$\sigma$) for the D image and a much larger offset of
0\farcs029 (15$\sigma$) for the main galaxy position.  The lens offset 
between the two reductions is about half an ACS pixel and may originate
with the different light profiles used when modeling the main lens galaxy 
(de Vaucouleurs + Exponential for our fit, single Sersic profile 
for the \citealt{astro-ph/0602309} fit).

\begin{figure}[t]
\plotone{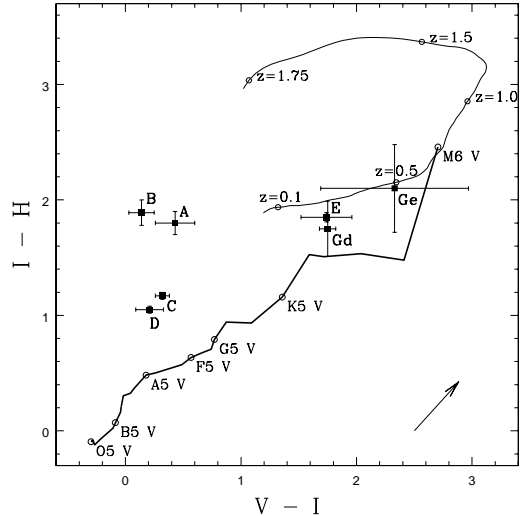}
\caption{Colors for \rxj\ system components A-E and lens galaxy components Gd
(de Vaucouleurs component) and Ge (exponential component).  Color tracks for 
the dwarf main sequence (heavy line) and a $z_f = 2.5$ burst galaxy model 
(thin line) are also drawn.  The arrow gives the reddening magnitude and 
direction assuming a $R_V=3.1$ Galactic extinction law and 
wavelength-dependent parameters from \citet{1989ApJ...345..245C}.}
\label{fig:colors}
\end{figure}

The colors of the system components are plotted in Figure~\ref{fig:colors}.  
Components A and B are three-quarters of a magnitude redder in $I-H$ compared 
to C and D, but all have nearly identical $V-I$ colors.  Component E is 
significantly redder by about $1.5$ 
mag in $V-I$ than the other four quasar images.  The reddening vector assuming
a $R_V = 3.1$ Galactic extinction law is given by the arrow in 
Figure~\ref{fig:colors}, and shows that E is unlikely a reddened copy of a 
fifth quasar image.  Its $I-H$ color is too red by several tenths of a 
magnitude for a late-K/early-M dwarf as well.  Its color is consistent with 
the primary lens galaxy, and \citet{astro-ph/0602309} suggested that $E$ may 
be an unresolved satellite galaxy.  

The main lens galaxy has total (exponential + de Vaucouleurs) colors 
in $V-I$ and $I-H$ of $1.84 \pm 0.11$ and
$1.83 \pm 0.20$, respectively.  These are bluer by $\sim$0.2 
magnitudes than expected for a $z_f = 2.5$ burst galaxy model observed at 
$z \approx 0.3$ (see Figure \ref{fig:colors}), but the difference is not 
unreasonable compared to the spread in galaxy colors observed in other lensed 
systems at similar redshifts \citep{2003ApJ...587..143R}.  As for the
exponential contribution to the galaxy model, its magnitude is fainter
than the de Vaucouleurs component by 2.1 mags, 1.5 mags and 1.2 mags in the 
$V$, $I$ and $H$ filters, respectively.  Its colors are redder than the 
de Vaucouleurs component, as expected for a bulge-like component to the
lens galaxy.   However, as already noted, the 
morphology is not as expected for a true bulge and the spatial offset between 
the ``bulge'' and ``disk'' components evident in Figure~\ref{fig:hstclose}e is 
puzzling.

\section{{\it Spitzer} Observations}\label{sec:spitzer}

We also observed J1131 with the Infrared Array Camera (IRAC) on board the 
{\it Spitzer} Space Telescope on 11 June 2005.  Observations consisted of 
a 36-point dither pattern for each of the four IRAC bandpasses (3.6, 4.5, 5.8,
and 8 $\mu m$), covering an area of roughly 7\farcm5 $\times$ 7\farcm5 
centered on the lens.  The data received from the Spitzer Science Center had 
passed through the standard IRAC pipeline (consisting of flat-field 
corrections, dark subtraction, and linearity and flux calibrations) and the 
resulting Basic Calibrated Data (BCD) files served as our input for MOPEX 
drizzling.  The BCD images were combined using a final scale of 0\farcs863 
pixel$^{-1}$, and Figure~\ref{fig:chandra} shows the [3.6] image of \rxjs\
and surrounding field.

Even after drizzling onto the finer pixel grid, the \rxjs\ A-D separation
spanned only four pixels.  The lack of spatial resolution complicates any 
analysis of the lens components and we instead focused on the IR colors 
of field galaxies.  After identifying extended sources in the field using the 
SExtractor package \citep[ver. 2.2.2]{1996A&AS..117..393B}, we searched 
for red sequences (e.g., \citealt{2000AJ....120.2148G}) in the IR 
color-magnitude diagrams indicative of overdense regions of elliptical and 
S0 galaxies (groups or clusters).  The resulting CMDs were suggestive
of a low-redshift ($z < 0.3$) overdensity of galaxies with [3.6]-[4.5] colors 
of $\sim0.1$, but the concentration was not defined enough to estimate a 
redshift.  Since the {\it Spitzer} observations were obtained, 
\citet{astro-ph/0511593} have indeed identified two red sequences 
for the \rxjs\ field using deep $VRI$ observations obtained with the KPNO 
Mayall 4m and CTIO Blanco 4m telescopes (see their Figure~8).  The 
dominant red sequence has a centroid (denoted by the eastern bullseye in 
Figure~\ref{fig:chandra}) located 2\farcm1 East of \rxjs\ G and 
$\approx$45\arcsec\ to the South-South-West from a bright elliptical galaxy 
identified in the mid-infrared images (G0 in Figure~\ref{fig:chandra}).  
Galaxy G0 is suggestive of a cD galaxy associated with the primary 
\citet{astro-ph/0511593} red sequence, implying significant structure along
the line of sight toward \rxjs\ (additional evidence supporting this claim 
is presented using the archival X-ray observations described below).  The
second, weaker red sequence detected by \citet{astro-ph/0511593} has a 
flux-weighted centroid 25\arcsec\ North-West of \rxjs\ G (western bullseye
in Figure~\ref{fig:chandra}) and includes the lens galaxy itself.

\section{{\it Chandra} Observations}\label{sec:chandra}

\begin{figure}[t]
\epsscale{1.15}
\plotone{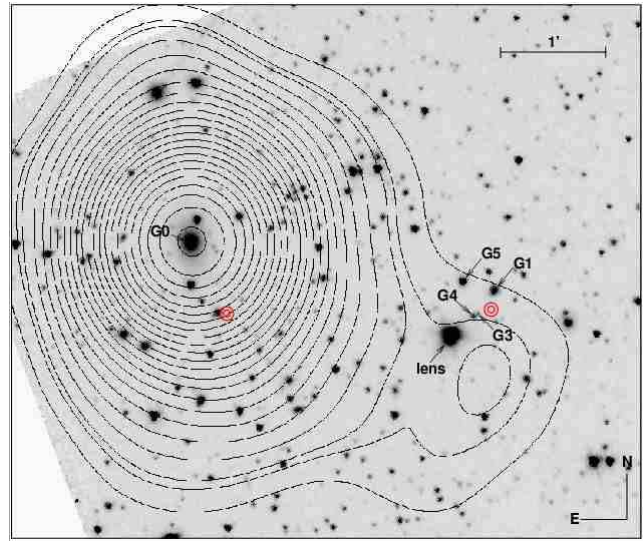}
\caption{Smoothed X-ray contours plotted on top of the IRAC Channel 1 (3.6 
$\mu$m) image.  X-ray contours are logarithmic except for the first 
two contour levels, which mark the $2\sigma$ and $3\sigma$ confidence levels.
The eastern (western) bullseye marks the raw centroid of the primary 
(secondary) galaxy red sequence detected by \cite{astro-ph/0511593}.}
\label{fig:chandra}
\end{figure}

Given the evidence for structure along the line of sight, we re-analyzed the 
archival 10~ks \chandra\ observation of \rxjs\ (see 
\citealt{astro-ph/0509027}) taken on 12 April 2004 using the ACIS-S3 
detector, with the goal of identifying X-ray emission from halos 
associated with the \citet{astro-ph/0511593} red sequences.  The smoothed 
X-ray emission (after excising the X-ray flux associated with the lensed 
quasar) is shown as contours on top of the IRAC [3.6] image in 
Figure~\ref{fig:chandra}.  We identified two extended sources of X-rays in the 
field aside from the lens flux analyzed by \citet{astro-ph/0509027}.  The 
first is the bright emission centered 153\arcsec\ North-East of 
\rxjs\ (at 11$^{\rm h}$32$^{\rm m}$1\fs4, $-12$\degr\ 31\arcmin\ 6\arcsec\, 
J2000; centroid accuracy of approximately 2\arcsec).  The X-ray position 
is coincident with the large elliptical galaxy (G0) identified in the 
mid-infrared images and within an arcminute of the red sequence flux 
centroid identified by \citet{astro-ph/0511593}.  This galaxy (and two others 
within three arcminutes) has a redshift measured from the Las Campanas 
Redshift Survey of $z=0.10$ and the region was identified as an 
optically-selected group by \citet{2000ApJS..130..237T}.  Thus, the red 
sequence, the cD galaxy and the extended X-ray emission confirm the presence 
of a foreground cluster at $z=0.1$.

We extracted the ACIS spectrum of the extended emission and used XSPEC 
\citep{1996ASPC..101...17A} to fit it with a thermal plasma model modified 
by Galactic absorption and obtained a good fit to the data with a $\chi^2=22$ 
for 41 degrees of freedom.  We estimated an unabsorbed 0.4-8 keV flux of 
$(7.5\pm0.8)\times10^{-13}$ \flux, a bolometric X-ray luminosity of 
$(3.1\pm0.5)\times10^{43}$ \lumin\ assuming the cluster is at $z=0.1$, and a 
temperature of $1.13\pm0.03$ keV.  The temperature and luminosity are roughly 
consistent with typical $L$-$T$ relations (e.g., 
\citealt{1999ApJ...524...22W, 2000MNRAS.319..933H, 2000ApJ...538...65X,
2002ARA&A..40..539R}) for groups and clusters, providing further confirmation 
of the cluster redshift.

The second extended source of X-ray emission is South-West of the lens, but 
it is fainter and the analysis is complicated by other point sources 
in the field.  To proceed, we did a general search for faint extended emission
following the procedures used in \citet{2005ApJ...625..633D}.  We used the 
CIAO tool \verb+wavdetect+ to identify point sources in the ACIS-S3 field and 
then replaced them with nearby background regions to create a soft band 
(0.5-2 keV) image.   We smoothed this image with a Gaussian of width 
$\sigma=25$\arcsec\ to find a source 33\arcsec\ Southwest of the lens
(11$^{\rm h}$31$^{\rm m}$50\fs1, $-12$\degr\ 32\arcmin\ 23\arcsec, J2000).  
Because of the low ($\sim 3\sigma$) significance of the detection and the 
effects of masking the overlapping emission from the lens, the position 
uncertainty is comparable to the smoothing scale.  Since this flux overlaps 
the lens and is close to the centroid of the weaker red galaxy sequence 
detected by Williams et al. 2005 (western bullseye in 
Figure~\ref{fig:chandra}),
we tentatively identify the emission as from the lens group at redshift of 
$z_l=0.295$.  The measured count rate in the 0.5-2~keV band is 
0.004 cts~s$^{-1}$ within a circle of radius 25\arcsec, corresponding to an 
unabsorbed flux of $1.7\times10^{-14}$ \flux.  If we use a $\beta$ model to 
extrapolate the X-ray flux to a larger aperture and assume a temperature of 
$T \simeq 1.5$~keV, then we estimate that the bolometric luminosity of this 
second cluster is $2\times10^{43}$ \lumin.  Thus, the halo associated with the
lens group is comparable in bolometric luminosity to the halo associated with
the low redshift group, but would require 10$\times$ the integration time to 
characterize at the same signal-to-noise.

\section{Lens Models and Interpretations}\label{sec:models}

In this section, we describe our mass modeling efforts aimed at reproducing 
the {\it HST} geometry and the SMARTS time delays.  We discuss three 
model variations.  First, we look at both isothermal 
and variable-slope power-law halo models for the main lens galaxy 
(\S\ref{sec:iso}), but in general find that such models cannot explain the 
long ($\sim$ 10 days) A--B and A--C time delays.  Next, we try a 
non-parametric mass model of the system (\S\ref{sec:nonpara}).  In general, 
such models add no information beyond that available in parametric models, 
but the asymmetric mass reconstruction suggests a satellite galaxy or other 
substructure near image A may be required to explain the anomalous 
delays.  Finally, we return to parametric models (\S\ref{sec:substruc}) 
consisting of the primary isothermal halo for the main lens and a secondary 
halo for the presumed substructure near image A.  Such models can plausibly 
reproduce both the long A delays and the image astrometry at the expense
of consuming the remaining degrees of freedom.

\subsection{Basic Isothermal Models}\label{sec:iso}

We first tried to model \rxjs\ using the standard singular isothermal 
ellipsoid embedded in an external shear \citep{1993ApJ...417..450K, 
1994A&A...284..285K}. For constraints, we used the {\it HST}/ACS $H$-band 
positions (Table~\ref{tab:astromphot}) with error bars of 3 mas for the quasar
positions and the three SMARTS time delays.  We looked at three model 
sequences:  model SISx uses a spherical halo centered on the lens galaxy 
plus an external shear; model SIEx allows for halo ellipticity plus shear; 
and model SIEx+ allows the halo center to move with respect to the galaxy 
center but constrained with errorbars of 5 mas.  The astrometry and time 
delays give 13 constraints, so the models have 6, 4 and 4 degrees of 
freedom, respectively.  We ignored the flux ratios.  All models were 
minimized in the image-plane using the {\tt gravlens} software of 
\citet{astro-ph/0102340} and adopting a $(\Omega_m, \Omega_{\Lambda}) 
= (0.3,0.7)$, $H_{\rm o} = 70$ km s$^{-1}$ Mpc$^{-1}$ cosmology.  The model 
results are summarized in Table~\ref{tab:models}.

\begin{figure}[b]
\plotone{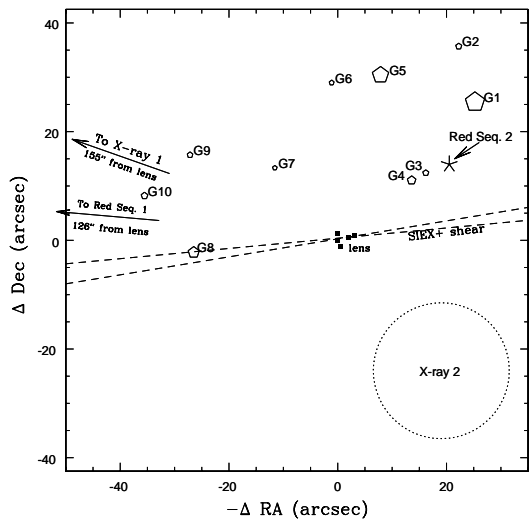}
\caption{Schematic of RXJ1131$-$1231 environment.  Labels X-ray 1 and 2 denote
the large and small X-ray emissions detected in the {\it Chandra} 
observations, and Red Seq. 1 and 2 denote the two red sequence centroids 
identified by \cite{astro-ph/0511593}; the dashed lines marks the 1$\sigma$
spread in the external shear direction for the SIEx+ model; the pentagons 
mark the locations of neighboring galaxies identified in the {\it HST}/ACS 
data.  The diameter of the X-ray 2 circle is the X-ray smoothing scale of 
25\arcsec.  Symbol sizes for the galaxies are proportional to the square root 
of their ACS $I$-band fluxes.  The lens galaxy $I$-band flux (not depicted by 
symbol size) is nearly twice that of G1.}
\label{fig:objects}
\end{figure}

None of the models gives a statistically acceptable fit to the observations; 
the best-fit SIEx+ model has a final $\chi^2$/dof of 74.1/4.  The 
simpler SISx and SIEx models cannot reproduce the image 
astrometry, with rms values between the observed and predicted quasar 
locations of 33 and 13 mas, respectively (final $\chi^2$/dof values of
538/6 and 138/4, with more than half of the $\chi^2$ budget coming from the
quasar positions).  The SIEx+ model fits the quasar positions
essentially perfectly (image rms of 2 mas) at the expense of shifting the 
halo center by 12 mas ($4\sigma$) from the measured luminous galaxy position.  
The situation is worse if we use the \citet{astro-ph/0602309} image and
galaxy positions, which yield $\chi^2$/dof values of 534/6, 158/4, and 130/4 
for the SISx, SIEx and SIEx+ models, respectively.  A substantial external 
shear is also required for both the Table~\ref{tab:astromphot} and 
\citet{astro-ph/0602309} positions -- greater than 10\% for all models -- 
and points at $-83$\degr\ East of North for the SIEx+ case (where the 
convention is for the shear direction to point towards or away from the 
perturber.)  There is no obvious perturber 
along the shear direction (see Figure \ref{fig:objects}); a galaxy with 
roughly the same velocity dispersion as the lens would produce a 10\% shear 
at 9\arcsec, but the nearest galaxies of significance (G3 and G4) are 
20\arcsec\ away and at the wrong position angle.  Galaxy G8 lies at the 
correct position angle, but is much too distant to produce a 10\% shear given 
its brightnesses and that of the main lens galaxy (see 
Figure~\ref{fig:hstwide}).  The shear does not point toward either of the 
X-ray halos detected in the {\it Chandra} data, or toward either of the red 
sequence centroids identified by \cite{astro-ph/0511593}.  This likely 
indicates that either there is no single external perturber dominating the 
shear or that the external shear is attempting to compensate for an inadequacy
in the main lens model.

The more troublesome problem is that all models fail to 
reproduce the measured time delays.  The SIEx+ profile predicts A--B and 
A--C delays of 0.98 and 1.23 days (assuming $H_{\rm o}$ = 70 km s$^{-1}$ 
Mpc$^{-1}$) compared to the measured values of roughly 12 and 10 days.  
Formally these are 9$\sigma$ and 5$\sigma$ discrepancies.  Note that the 
predicted B--C delay of 0.25 days for the SIEx+ model agrees within 
2$\sigma$ with the measured value of \delayBConesig\ days, so the problem 
is chiefly with the A image. The situation cannot be corrected by simply 
rescaling $H_{\rm o}$, since matching the A--B and A--C delays would require 
an unrealistic Hubble constant of $\approx$ 10 km~s$^{-1}$ Mpc$^{-1}$.

The long A delay is difficult to explain.  
There are many examples of model perturbations leading to systematic 
uncertainties in time delay predictions (see \citealt{2005IAUS..225..281S} 
for a review), but these typically alter the delay predictions at the 10\% 
level and not the factor of 10 needed here.  For example, the mass-sheet 
degeneracy \citep{1988ApJ...327..693G} rescales the image delays by 
$\Delta \tau \propto (1 - \langle \kappa \rangle)$ by superimposing a 
mass-sheet of surface density $\langle \kappa \rangle$.  One might 
expect a convergence of $\langle \kappa \rangle \simeq 0.1$-$0.3$ from matter 
associated with the nearby X-ray halos (e.g., as with systems Q0957+561, 
\citealt{1999ApJ...520..479B}; RXJ 0911+0554, \citealt{1998ApJ...501L...5B}),
but this would uniformly shorten all delays by 10\%-30\%, 
and not preferentially lengthen the A delay.  To a similar degree, small 
variations in the radial exponent of the halo density profile ($\sim$ 10\%) 
alter delay predictions by a comparable fraction \citep{2002ApJ...578...25K}, 
again of insufficient magnitude to explain the A delay.  An example of an 
extreme change in the density slope is to use a de Vaucouleurs constant 
mass-to-light ratio model with an effective radius matching the value measured
from the NICMOS data.  Such a model can match the astrometry as well as the 
SIEx+ model provided the de Vaucouleurs center is allowed to float, but the
predicted cusp delays are still $\sim 1$ day.  

Abandoning the assumption of isothermality, we also explored a range of 
elliptical power-law profiles with three-dimensional density profiles falling
as shallow as $1/r$ to as steep as $1/r^4$.  We found that highly elliptical
profiles steeper than $1/r^3$ could produce long cusp delays provided
the halo was orientated along the A-G line.  Because of the high ellipticity, 
such models concentrated more mass near A than near B and C, increasing the 
Shapiro contribution to the A delay.  However, the required ellipticity to 
produce an A--B delay of even half the observed value was flatter than 
$e = 0.9$.  Such a mass distribution would indicate a prominent disk galaxy, 
and this is clearly inconsistent with the morphology of the lens observed in 
the NICMOS data.

\subsection{Non-Parametric Models\label{sec:nonpara}}

\begin{figure}[t]
\plotone{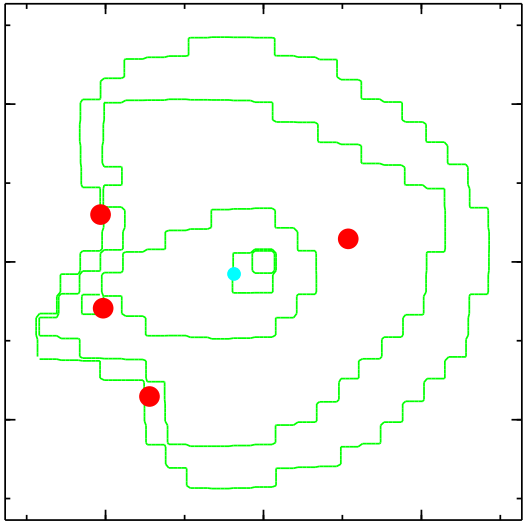}
\caption{PixeLens ensemble-average mass map of \rxj, as constrained by the
system astrometry and ABC time delays.  The mass contours are logarithmically 
spaced at factors of 2.5 in surface density, beginning with the outer contour 
at $\kappa \approx 0.16$.}
\label{fig:pixelens}
\end{figure}

Lacking a successful macromodel, we turned to non-parametric models using the 
PixeLens software of \citet{2004AJ....127.2604S}.  As noted above, such models
do not add any more information than present in parametric models since 
they are grossly underconstrained, but they may indicate where the parametric
models are qualitatively wrong.  PixeLens works by 
reconstructing a pixelated mass map of the lens using the image positions 
and time delays as constraints.  For \rxjs, we reconstructed both symmetric 
and asymmetric mass models over a 6\arcsec\ $\times$ 6\arcsec\ grid centered 
on the lens with a ``pixrad'' of 11, yielding around 530 separate mass 
elements.  Since PixeLens assumes negligible errors for all input values, we 
tried models both with and without the more uncertain D delay but found 
little qualitative change in the final mass map.  

The reconstruction for the 
asymmetric mass distribution is shown in Figure~\ref{fig:pixelens}.  
(For the symmetric reconstruction, the eastern half of the structure 
is mirror-imaged on the western half, producing an unrealistic four-leaf 
clover mass distribution). 
There is a clear protrusion surrounding image A.  
PixeLens's solution for increasing the A delay is to have significant 
variation in the angular distribution of matter along the three cusp images 
by increasing the surface mass density around image A ($\kappa \simeq 0.9$) 
as compared to the surface mass densities at images B and C ($\kappa \simeq 
0.1$-$0.2$).  Qualitatively, this is similar to our result found for a highly 
elliptical disk-like halo overlapping the A image.  However, the asymmetric 
reconstruction stresses that what is chiefly needed is a higher surface mass 
density at A compared to the other two cusp images, suggesting that a 
satellite galaxy or other undetected substructure at this location may be 
responsible for the anomalous delay.

\subsection{Substructure Near Image A}\label{sec:substruc}

There is no visual sign of a satellite galaxy near image A in the {\it HST}
images or the corresponding subtraction residuals.  A red satellite 
galaxy would be most evident in the F160W image, but the significant 
Einstein ring emission and PSF residuals complicates any visual
search near the lensed images.  To explore if a perturber 
could reproduce the cusp delays, we added an SIS mass model over a 
grid of positions surrounding image A and looked at how each model 
adjusted for the \rxjs\ time delays.  The main lens galaxy was again modeled 
using the SIEx+ profile, and we fitted both the astrometric and delay 
constraints as before.   We found that a perturber with an 
Einstein radius of $\sim0\farcs2$ and placed within $\sim0\farcs1$ South-East 
of the A image could lengthen the A--B and A--C delays by several days, 
but such models 
consistently formed two additional images of the background 
quasar.  The perturber's proximity to image A means that its critical curve 
always maps into a source-plane caustic near the source position, creating a 
six-image region inside the primary lens astroid caustic.  While one of these 
extra images is always highly demagnified and located at the perturber 
position, the other typically forms at twice the perturber-A separation with 
a magnification $\sim 10\%$ that of image A.  Such an image would be easily 
identified in the {\it HST} data.

We can avoid the formation of extra images by adding a core radius
$s$ to the perturber's profile, such that the projected density
is given by
\begin{equation} 
\kappa(r) = \frac{b'}{2}\frac{1}{\sqrt{s^2 + r^2}} 
\end{equation}
(e.g., \citealt{1993ApJ...417..450K, 1994A&A...284..285K}).  The exact 
criteria needed to prevent extra images depend on the perturber's core size 
and mass scale.  Consider a model where the main lens (modeled as an SIE)
and perturber (modeled as a softened isothermal sphere, or IS) are colinear 
along the x-axis with the main lens at the origin and the perturber  
at a distance $x=x'$.  Including an external shear term (with shear angle 
$\theta_{\gamma}$ measured from the x-axis), the potential 
$\phi(x)$ is
\begin{equation}
\phi(x) = \phi_{\rm SIE}(x) + \phi_{\rm IS}(x-x') + \frac{\gamma x^2}{2}\cos(2\theta_{\gamma}).
\end{equation}
The appearance of extra images first becomes possible (that is, a critical 
line and corresponding six-image caustic are first formed) when a solution to 
the lens equation exists at $x=x'$.  This translates into a condition on the 
potential of 
\begin{equation}
\frac{\partial^2 \phi}{\partial x^2}\arrowvert_{x=x'} = 1,
\end{equation}
or, after substitution, 
\begin{equation}
\frac{b'}{2s} \approx 1,
\end{equation}
where we have neglected a shear term that is always much smaller than unity 
for reasonable models.  No extra images are possible when $b'/2s \lesssim 1$ 
and multiple images are possible (provided the source is inside the six-image 
caustic) otherwise.  

\begin{figure}[t]
\plotone{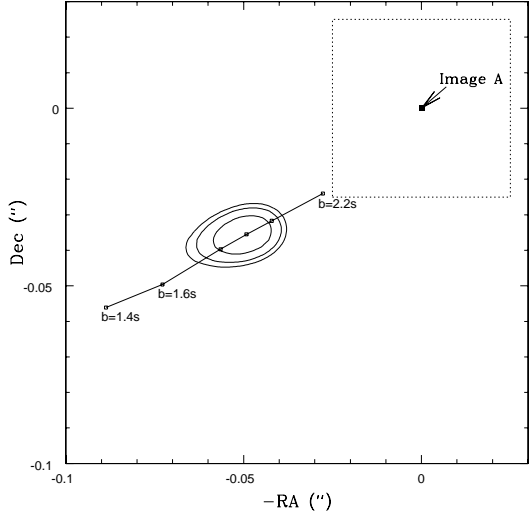}
\caption{Optimal perturber positions for the best-fit sub-critical 
($b=1.4s, b=1.6s, b=1.8s$), critical ($b=2.0s$) and super-critical 
($b=2.1s, b=2.2s$) models.  The three contours outline the 
$\Delta \chi^2 = 2.6, 4.3, 6.2$ ($1\sigma$, $2\sigma$, and 3$\sigma$ for 
two parameters) contour parameters for the critical case.  The location of 
image A is marked.  For scale, the dotted box shows the size of one ACS 
pixel centered on image A.}
\label{fig:ispos}
\end{figure}

\begin{figure}[b]
\plotone{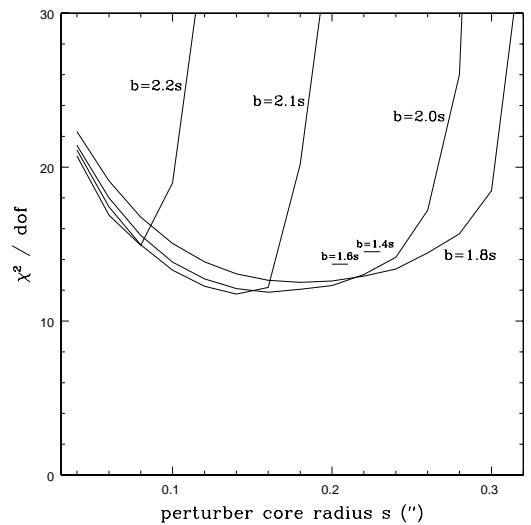}
\caption{Goodness of fit for two super-critical ($b=2.2s$ and $b=2.1s$),
critical ($b=2.0s$) and sub-critical ($b=1.8s$) perturbers near image A.
Minimums for the $b=1.6s$ and $b=1.4s$ cases are also noted by the two tick
marks.}
\label{fig:critical}
\end{figure}

We looked at a range of perturber models: two super-critical ($b=2.2s$ 
and $b=2.1s$), a critical ($b=2.0s$) and three sub-critical ($b=1.8s$, 
$b=1.6s$ and $b=1.4s$).  During the minimization, a large penalty term was 
added to the merit function whenever the source position crossed into the 
six-region caustic, insuring that only four-image models were accepted.  We 
also constrained the orientation of the main lens halo using the 
measured NICMOS position and associated error, which helped stabilize the 
ellipticity of the primary lens.  This approach gave a final tally of 12 
parameters (two for the source position, seven for the main lens, and three 
for the perturber) and 14 constraints for two degrees of freedom.  
Table~\ref{tab:models} summarizes the result for the critical perturber.  

Figure~\ref{fig:ispos} shows the optimal perturber positions for the 
model sequence and the confidence contours for the critical 
($b=2.0s$) case.  For all models, the perturber position was tightly 
constrained to a region between $0\farcs03$ to 0\farcs11 South-East of 
image A.  In Figure~\ref{fig:critical}, we show the resulting $\chi^2/dof$ 
for each model as the perturber's core radius is increased from $s=0$\arcsec\ 
(the singular case) to $s=0\farcs3$ (about 1.3 kpc at the lens redshift).  
Note that for the two super-critical models, $\chi^2$ rapidly increases once a 
threshold core radius is crossed, with roughly 90\% of the increase coming 
from the quasar positions.  The abrupt increase is due to the 
growing size of the six-image caustic as the perturber's core radius and 
potential depth are increased, which displaces the source position from its 
optimal location inside the six-image region.  This means a super-critical 
perturber is highly unlikely since it requires a fine-tuned source position 
in order to match the quasar astrometry while remaining outside the six-image 
region.  The critical and sub-critical models avoid the formation of the 
six-image caustic and consequentially show no abrupt increase in $\chi^2$.  

As seen from Figure~\ref{fig:critical}, the critical model is marginally 
preferred over the sub-critical cases.  There are two notable improvements
for the optimal critical model compared to the single halo model considered in 
\S\ref{sec:iso}.  First, both the quasars and galaxy astrometry are now fitted
essentially perfectly: $\chi^2/dof$ = 0.3/2 compared to 13/4 for the SIEx+ 
case, with virtually all (99\%) of the $\chi^2$ budget now coming from the 
time delays.  Second, the cusp delays are lengthened considerably:  A--B, 
A--C and A--D delays of 5.90, 7.88 and $-80$ days, respectively.  The 
remaining model parameters are quite reasonable as well, with a nearly 
round main lens galaxy (ellipticity of 0.14) and small external shear (4\%).  
Thus, it is possible to improve both the overall astrometry and the time delay
predictions by adding a perturber halo nearly coincident with image A, 
provided the halo is soft enough to avoid the formation of extra images.
 
The A--B delay is still a factor of two smaller than measured (a 4$\sigma$ 
discrepancy).  Since the Shapiro contribution to the time delays depends on 
differences in effective potential between images, one simple way to fine-tune
the delays is to vary the radial slope of the perturber.  With a variable slope
$\alpha$, the projected surface density $\kappa(r)$ can be written as
\begin{equation}
\kappa(r) = \frac{1}{2}\frac{b'^{2-\alpha}}{({s^2 + r^2})^{1-\alpha/2}},
\label{equ:powerlaw}
\end{equation}
where again $s$ is the core radius and the normalization is chosen such that 
$b$ is the system's Einstein radius in the limit of an isothermal profile 
($\alpha \rightarrow 1$) and vanishing core radius.  For small core radii, 
the effective potential is approximately given by $\phi(r) 
\approx b'^{2-\alpha} r^{\alpha} / \alpha^2$.  Generating longer A--B and A--C 
delays requires a larger difference in the A--B and A--C effective potentials,
which is possible for steeper than isothermal ($\alpha<1$) mass profiles.

To test if a non-isothermal slope could more closely match the delays,
we allowed the potential depth $b'$ and halo slope $\alpha$ to vary 
independently while again looking at a range of core radii $s$.  This 
modification consumes the remaining two degrees of freedom and we expect a 
perfect fit for a plausible model.  Indeed, we found that a halo slope 
with $\alpha = 0.7$-$0.8$ could reproduce both the astrometry and measured
time delays essentially perfectly although with a clear degeneracy
between the size of the core radius and the halo slope.  For the 
$s=0\farcs2$ model (summarized in Table~\ref{tab:models}), the predicted
delays were 12.0, 13.8 and -85 days for A--B, A--C and A--D, respectively,
with an overall $\chi^2/dof$ of 4.7/0.  It is interesting that we still find 
$\chi^2 > 0$ even for zero degrees of freedom.  The only $\chi^2$ 
contribution for all $s \gtrsim 0\farcs15$ models is from the A--C delay,
although it is more instructive to look at the discrepancy using the B--C 
difference.  The predicted B--C delays (between +1.0 and +1.8 days for the
range of core radii considered) is in the opposite sense of the
measured value (\delayBConesig\ days).  The delay ordering among images 
is a model-independent feature of a lens (e.g., 
\citealt{2003AJ....125.2769S}), so if 
the sense of the delay is measured incorrectly then even a model with zero
degrees of freedom will be unable to reproduce the incorrect delay sense.  All
models considered in this paper predict B trailing C (B--C $> 0$), 
implying that images A and B are the ``merging'' image pair normally seen
in inclined-quad configurations and that the overall image ordering is 
CBAD.  Therefore, if the measured C delay is off by 2-3 $\sigma$ such that the
measured B--C delay becomes positive, then it might explain the 
non-zero $\chi^2$.

\begin{figure}[t]
\plotone{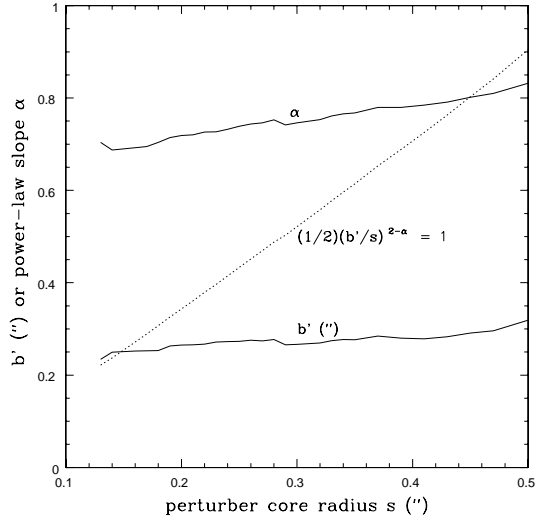}
\caption{Potential strength ($b'$) and halo slope ($\alpha$)
for a non-isothermal perturber.  The dotted line marks the potential
strength for the just-critical perturber, using each model's core radius
and best-fit halo slope.}
\label{fig:isalpha}
\end{figure}

Figure~\ref{fig:isalpha} shows the degeneracy between the perturber slope
$\alpha$ and core radii $s$, as all models in the figure yield $\chi^2/dof$
values between 3/0 and 5/0.  We also plot the best-fit potential 
depth $b'$ for the perturber and the condition for the just-critical 
case (dotted line), which for the power-law profile is 
\begin{equation}
\frac{1}{2}\left(\frac{b'}{s}\right)^{2-\alpha} \approx 1.
\end{equation}
The total
$\chi^2/dof$ gradually drops from $4.7/0$ at $s=0\farcs15$ to $3.0/0$ at
$s=0\farcs45$ with again the only significant contribution coming from the C 
delay.  Over this range, $b'$ if fairly constant, ranging from 0\farcs25 to 
0\farcs28, and the halo slope increases from $\alpha$ = 0.7 to 0.8.  
Note that for $s\gtrsim0\farcs15$, the perturber no longer sits on the 
just-critical line as seen for the optimal ($b=2s$) isothermal model.  Thus
a steeper-than-isothermal profile not only reproduces the long A delay
but also naturally avoids the formation of extra images.

\subsection{Mass of the Perturber}

The cylindrical mass for the surface density given in Eq.~\ref{equ:powerlaw} 
increases as $M_{\rm cyl}(r) \propto r^{\alpha}$, so formally the total 
perturber mass in the adopted model is unbounded.  Adopting
a cut radius comparable to the A-C or A-B image separations ($\approx 
$1\farcs2, or 5.3 
kpc at the lens redshift) gives an enclosed perturber mass of $4.7 \times 
10^{10}$ M$_{\sun}$.  As a comparison, the mass enclosed by the primary lens 
galaxy out to the same cylindrical radius (centered on its respective 
profile) is $4.4 \times 10^{11}$ M$_{\sun}$, so the perturber has $\lesssim 
10\%$ of the mass of the main lens at comparable radii.  This is fairly 
significant and one might expect some visual evidence if it is indeed a 
satellite galaxy.  We can estimate a plausible luminosity to the perturber by 
scaling its velocity dispersion with that of the main lens and 
using the \citet{1976ApJ...204..668F} relationship.  For an enclosed mass 
that scales with velocity dispersion as $M(R) \propto \sigma^2$, the factor 
of 10 difference in mass translates into a factor of 3.2 in velocity 
dispersion, or a factor of 100 difference in luminosity assuming 
$L \propto \sigma^4$.  Such a galaxy would be extremely difficult to see 
given the ring emission and light from image A, even after image subtraction.
For example, scaling from the observed count-rate of the main lens measured 
in the NICMOS image, the predicted {\it peak} count rate from the perturber 
is only $\sim$0.008 cnts s$^{-1}$.  The rms count rate within a 0\farcs4 
radius around image A after subtracting the quasar images and ring emission 
(Figure~\ref{fig:hstclose}f) is 0.058 cnts s$^{-1}$, which is larger than
the peak signal expected from the satellite galaxy by a factor of seven.  Thus
even if the perturber were a {\it bona fide} satellite galaxy, it is doubtful
we would detect it in the existing {\it HST} images.

\subsection{Placement on the Fundamental Plane}

One check on the models is to compare the predicted velocity dispersion of 
the main lens galaxy with the value expected from the Fundamental Plane (FP; 
\citealt{1987ApJ...313...59D, 1987ApJ...317....1D}).  Almost all lens 
galaxies lie on the present-day FP provided one allows for passive luminosity 
evolution with redshift \citep{2001MNRAS.326..237T, 2002ApJ...564L..13T, 
2003ApJ...587..143R}.  When studying gravitational lenses, it is convenient 
to phrase this evolution in terms of the galaxy's $B$-band mass-to-light 
ratio $(M/L)_B$ measured inside the system's Einstein radius (see 
\citealt{2001MNRAS.326..237T} for a thorough discussion).  Parameterizing to 
first-order in redshift as $\Delta \log (M/L)_B = az$, studies of 
gravitational lens galaxies find values between $a \approx 0.5$ 
\citep{2003ApJ...587..143R} and $a \approx 0.7$ \citep{astro-ph/0512044}.  
Similar results are found for cluster \citep{1998ApJ...504L..17V} and field 
\citep{2001ApJ...553L..39V} galaxies out to $z \approx 0.8$ that are 
not associated with lensing.

\begin{figure}[t]
\plotone{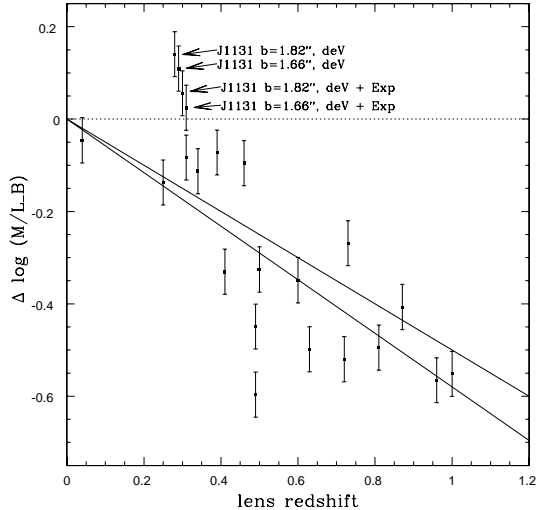}
\caption{$M/L_B$ evolution of the J1131 lens galaxy (arrows offset in 
redshift for clarity) from the present-day FP, computed using combinations of 
the total de Vaucouleurs plus Exponential galaxy magnitude, just the de 
Vaucouleurs component magnitude, and the two velocity dispersion estimates 
described in the text.  The other 18 points are evolutions obtained for lens 
galaxies listed in \citet{2003ApJ...587..143R} that have spectroscopically 
measured lens redshifts.  The solid lines denote the best-fit 
\citet{2003ApJ...587..143R} slope of $\Delta \log(M/L) = -0.54 \pm 0.04$ using
28 gravitational lens galaxies.}
\label{fig:dellogml}
\end{figure}

Using a $z_f = 2.5$ starburst SED to calculate the necessary 
$k$-corrections and evolution effects, the restframe $B$-band 
surface brightness for \rxjs\ G is 20.62 $\pm$ 0.05 mag arcsec$^{-2}$ using 
the total (de Vaucouleurs plus exponential) galaxy magnitude and 20.83 $\pm$ 
0.03 mag arcsec$^{-2}$ using just the de Vaucouleurs galaxy magnitude. 
The system's velocity dispersion is either 353 $\pm$ 2 km s$^{-1}$ or 337 
$\pm$ 2 km $s^{-1}$ for the SIEx+ or SIEx+ plus $\alpha=0.72$ perturber 
model.  Both values
correspond to extremely bright galaxies, $6L_*$ and $5L_*$ respectively for 
$\sigma_* = 225$ km~$s^{-1}$.  Such galaxies are rare but not
unheard of.  The \citet{2003AJ....125.1817B} SDSS sample of $\sim$9000 field 
ellipticals contains galaxies with velocity dispersions as large as 400 
km~s$^{-1}$.  Although a $\sigma = 350$ km~s$^{-1}$ galaxy is 200 times less 
common than a $\sigma \approx 170$ km~s$^{-1}$ galaxy found at the peak of 
the elliptical velocity dispersion function \citep{2003ApJ...594..225S},
the cross-section for lensing scales as $\sigma^4$ so massive galaxies are 
preferentially favored.  Finally, the galaxy intermediate-axis effective 
radius measured from the NICMOS data gives a physical size of $R_e = 5.0 
\pm 1.2$ kpc.

The evolution rates required to place \rxjs\ G onto the present-day FP are
plotted in Figure~\ref{fig:dellogml}.  Four estimates are computed, using
the possible combinations of the two surface brightness estimates and the two
velocity dispersion estimates.  Also plotted are rates computed for 18 other 
gravitational lenses with measured lens and source redshifts listed in 
\citet{2003ApJ...587..143R}.  Contrary to the \citet{2003ApJ...587..143R} 
sample of lens galaxies, all \rxjs\ G evolution rates are inconsistent 
with passive luminosity evolution, with only the halo plus perturber
model and total (de Vaucouleurs plus exponential) galaxy magnitude rate 
consistent with no evolution at all at the 1$\sigma$ level ($a = 0.02 
\pm 0.05$).  Another way to view this is that the galaxy's inferred 
velocity dispersion is simply too large for its measured luminosity.  
Adopting $M_* = -19.79 + 5\log(h) \pm 0.04$ \citep{2002MNRAS.333..133M}, 
$\gamma = 4.0 \pm 0.25$ \citep{2003AJ....125.1849B} and $\sigma_* = 225 
\pm 22.5$ km~s$^{-1}$, the Faber-Jackson relationship yields $\sigma = 274 
\pm 28$ km~s$^{-1}$ using the combined de Vaucouleurs plus exponential 
magnitude.  This estimate is considerable lower than the 
value of $\sigma = 337 \pm 2$ km~s$^{-1}$ obtained from the halo plus 
perturber model, suggesting that there may be an additional source of mass 
contributing convergence inside the Einstein ring besides the main lensing
galaxy.  Part of the solution likely lies in the mass convergence associated
with the overlapping X-ray halos detected in the {\it Chandra} data and
the unknown object component E, if it is truly a satellite galaxy to the 
primary lens.  Both effects would lower the inferred value of the main lens 
velocity dispersion by accounting for some of the inferred mass inside the
system's Einstein ring.  

\section{Summary and Conclusions}\label{sec:conclusions}

We have measured the time delays for the quadruple lens \rxj\ based on two 
seasons of monitoring data from the SMARTS/CTIO 1.3m telescope.  The short
delays for the cusp images are A--B = \delayABonesig\ and 
A--C = \delayAConesig\ and the long delay for the counter image is A--D 
= \delayADonesig.  The A--B and A--C delays are an order of magnitude larger 
than expected using a standard elliptical model of the lens galaxy.
One plausible explanation arises from a satellite galaxy or other dark
substructure about $0\farcs05$ from the central cusp image,
although some fine-tuning is required to match the observed properties of 
the system.  In particular, for the power-law mass models explored here,
the perturber must have a core radius of $\gtrsim 0\farcs2$ ($\gtrsim$ 
1 kpc) to avoid the formation of extra images of the background quasar, a 
steeper than isothermal mass profile ($\alpha \approx 0.7; \rho 
\propto r^{-\alpha}$), and a significant total mass within the A-C and 
A-B image separations of $\sim 4.7 \times 10^{10}$ M$_{\sun}$ (about 10\% 
of the main lens galaxy for comparable radii).  Although massive, such a 
satellite galaxy would have a luminosity only $\sim$1\% of the main lens, 
making it undetectable in the extant data given the perturber's proximity
to the bright quasar A image and the prominent Einstein ring emission.

Ancillary evidence for such a perturber comes from the improved astrometry of
the quasar positions and primary lens galaxy.  As noted by 
\citet{astro-ph/0602309}, using a standard isothermal halo for the main lens 
can reproduce the relative quasar positions but not the position of the lensing
galaxy.  This was addressed by \cite{astro-ph/0602309} by adding an $m=4$
octupole term (characteristic of boxy/disky isodensity contours) to 
the lensing potential, although they admit that the physical nature of the 
term is difficult to interpret.  Comparable deviations from pure elliptical 
models have been ruled out for four other lens galaxies using the shapes of 
Einstein rings \citep{astro-ph/0511001}.  Here, we have shown that 
including a perturber near image A allows for an essentially perfect fit 
to the quasars and galaxy astrometry while simultaneously explaining the 
anomalous cusp delays.

We have also re-analyzed the {\it Chandra} X-ray data first presented by 
\citet{astro-ph/0509027} and found evidence for two X-ray halos along the
line of sight toward \rxjs.  The brighter halo is coincident with a foreground
($z=0.1$) cD galaxy 2\farcm5 North-East of the lens and corresponds to the
stronger of two galaxy red sequences detected in the optical by 
\citet{astro-ph/0511593}.  The fainter halo is much closer to the lens and
presumably includes \rxjs\ itself, given the halo's proximity to the lens 
and proximity to the weaker of the two galaxy red-sequences identified by 
\citet{astro-ph/0511593}.  

Overall, the data suggest a very complicated lensing environment for \rxjs.
The lens galaxy is likely part of a $z\approx0.3$ group or cluster
of galaxies with sufficient mass to support a $\sim1.5$ keV X-ray halo;
there is a foreground cluster of galaxies along the line of sight 
as evidenced by a second X-ray halo of similar temperature and centered on
a prominent $z=0.1$ cD galaxy; a significant mass perturbation near the A 
image is required to explain the anomalous time delays for the cusp 
images; modeling the surface brightness profile of the main lens requires at 
least two distinct profile shapes, suggestive of a more complicated primary
lens than a simple de Vaucouleurs-shaped elliptical; and finally, there is a 
faint, point-like object 0\farcs5 North of \rxjs\ G with colors similar to 
those of the main lens galaxy but otherwise of an as-of-yet undetermined 
nature.  These complications make it difficult at present to use the measured 
time delays to either estimate the Hubble constant or to constrain the overall
mass profile of the main lens.  The monitoring program has revealed extensive 
($\sim$ 0.3-0.4 mag) microlensing variability in the optical on time scales
of months, so future microlensing studies of the system are promising.

Progress with future mass modeling will at least require measuring the central
velocity dispersion of \rxjs\ G to break the mass-sheet degeneracy between the
main lens galaxy and X-ray halos identified in the {\it Chandra} data.
Such information will likely help to reconcile the lack of apparent $M/L$ 
evolution for the main lens as well, since it will provide an estimate of 
the halo convergence inside the Einstein ring that is currently assigned to 
the lens galaxy.  Deep imaging and spectral observations of component E are 
obviously required as well, but may prove challenging given its overall 
faintness, extremely red color, and proximity to the main lens.

\acknowledgments

This work is based in part on observations made with the NASA/ESA {\it Hubble Space Telescope}, obtained at the Space Telescope Science Institute, which is operated by AURA, Inc., under NASA contract NAS5-26555.  This research is supported by {\it HST} grants G0-9375 and GO-9744.  This work is also based in part on observations made with the Spitzer Space Telescope, which is operated by the Jet Propulsion Laboratory, California Institute of Technology under a contract with NASA. Support for this work was provided by NASA through an award issued by
JPL/Caltech.  These observations are associated with Spitzer program 20451.

\bibliography{ads}

\ \\
\ \\
\ \\
\begin{longtable}{cccccccc}
\caption[Lightcurves for RXJ~1131$-$1231]{Lightcurves for RXJ~1131$-$1231}\label{tab:lcurve} \\

\multicolumn{8}{c}{} \\ [-1.5ex]
\hline \hline \\[-2ex] 
       \multicolumn{1}{c}{HJD} &
       \multicolumn{1}{c}{$\chi^2$/dof} &
       \multicolumn{1}{c}{\phantom{++}Comp. A\phantom{++}} &
       \multicolumn{1}{c}{\phantom{++}Comp. B\phantom{++}} &
       \multicolumn{1}{c}{\phantom{++}Comp. C\phantom{++}} &
       \multicolumn{1}{c}{\phantom{++}Comp. D\phantom{++}} &
       \multicolumn{1}{c}{\phantom{++}ref. stars\phantom{++}} &
       \multicolumn{1}{c}{\phantom{++}Telescope\phantom{++}} 
        \\[0.5ex] \\[-3.2ex]
\endfirsthead

\multicolumn{3}{c}{{\tablename} \thetable{} -- Continued} \\[0.5ex]
  \hline \hline \\[-2ex]
   \multicolumn{1}{c}{HJD} &
   \multicolumn{1}{c}{$\chi^2$/dof} &
   \multicolumn{1}{c}{\phantom{++}Comp. A\phantom{++}} &
   \multicolumn{1}{c}{\phantom{++}Comp. B\phantom{++}} &
   \multicolumn{1}{c}{\phantom{++}Comp. C\phantom{++}} &
   \multicolumn{1}{c}{\phantom{++}Comp. D\phantom{++}} &
   \multicolumn{1}{c}{\phantom{++}ref. stars\phantom{++}} &
   \multicolumn{1}{c}{\phantom{++}Telescope\phantom{++}} 
        \\[0.5ex] \hline  \\[-1.8ex]
\endhead

  \multicolumn{3}{l}{{Continued on next page\ldots}} \\
\endfoot

 \hline
\endlastfoot
\\[-1.8ex] \hline

$2988.811$ &$  1.29$ &$ 2.379\pm 0.009$ &$ 2.233\pm 0.009$ &$ 3.192\pm 0.014$ &$ 4.560\pm 0.037$ &$\phantom{-} 0.006\pm 0.002$ &SMARTS \\ 
$2995.804$ &$  5.62$ &$ 2.465\pm 0.010$ &$ 2.220\pm 0.009$ &$ 3.088\pm 0.013$ &$ 4.380\pm 0.027$ &$\phantom{-} 0.009\pm 0.002$ &SMARTS \\ 
$3002.819$ &$ 10.58$ &$ 2.383\pm 0.007$ &$ 2.321\pm 0.007$ &$ 3.147\pm 0.009$ &$ 4.495\pm 0.021$ &$\phantom{-} 0.008\pm 0.002$ &SMARTS \\ 
$3013.819$ &$  2.33$ &$ 2.398\pm 0.008$ &$ 2.388\pm 0.008$ &$ 3.270\pm 0.013$ &$ 4.595\pm 0.034$ &$\phantom{-} 0.007\pm 0.002$ &SMARTS \\ 
$3021.790$ &$  6.08$ &$ 2.495\pm 0.009$ &$ 2.342\pm 0.009$ &$ 3.144\pm 0.013$ &$ 4.499\pm 0.028$ &$\phantom{-} 0.009\pm 0.002$ &SMARTS \\ 
$3028.765$ &$  4.21$ &$ 2.460\pm 0.007$ &$ 2.369\pm 0.007$ &$ 3.271\pm 0.010$ &$ 4.496\pm 0.021$ &$\phantom{-} 0.007\pm 0.002$ &SMARTS \\ 
$3035.815$ &$  3.93$ &$ 2.454\pm 0.007$ &$ 2.385\pm 0.007$ &$ 3.284\pm 0.010$ &$ 4.512\pm 0.021$ &$\phantom{-} 0.007\pm 0.002$ &SMARTS \\ 
$3039.758$ &$  2.89$ &$ 2.461\pm 0.008$ &$ 2.404\pm 0.008$ &$ 3.232\pm 0.012$ &$ 4.639\pm 0.035$ &$\phantom{-} 0.008\pm 0.002$ &SMARTS \\ 
$3042.783$ &$  0.76$ &$ 2.479\pm 0.012$ &$ 2.334\pm 0.011$ &$ 3.264\pm 0.021$ &$ 4.591\pm 0.059$ &$-0.003\pm 0.002$ &SMARTS \\ 
$3046.734$ &$  1.41$ &$ 2.440\pm 0.009$ &$ 2.361\pm 0.009$ &$ 3.249\pm 0.015$ &$ 4.547\pm 0.040$ &$\phantom{-} 0.004\pm 0.002$ &SMARTS \\ 
$3050.769$ &$  7.66$ &$ 2.443\pm 0.007$ &$ 2.361\pm 0.006$ &$ 3.216\pm 0.009$ &$ 4.547\pm 0.020$ &$\phantom{-} 0.007\pm 0.002$ &SMARTS \\ 
$3052.918$ &$  1.50$ &$ 2.505\pm 0.006$ &$ 2.365\pm 0.006$ &$ 3.341\pm 0.008$ &$ 4.740\pm 0.020$ &$\phantom{-} 0.008\pm 0.002$ &  APO \\ 
$3055.713$ &$  2.08$ &$ 2.491\pm 0.009$ &$ 2.328\pm 0.008$ &$ 3.213\pm 0.012$ &$ 4.555\pm 0.027$ &$\phantom{-} 0.008\pm 0.002$ &SMARTS \\ 
$3058.707$ &$  4.69$ &$ 2.457\pm 0.007$ &$ 2.324\pm 0.007$ &$ 3.271\pm 0.010$ &$ 4.567\pm 0.021$ &$\phantom{-} 0.006\pm 0.002$ &SMARTS \\ 
$3063.712$ &$  4.85$ &$ 2.466\pm 0.008$ &$ 2.377\pm 0.007$ &$ 3.271\pm 0.011$ &$ 4.574\pm 0.026$ &$\phantom{-} 0.006\pm 0.002$ &SMARTS \\ 
$3066.739$ &$  1.14$ &$ 2.518\pm 0.011$ &$ 2.322\pm 0.010$ &$ 3.243\pm 0.015$ &$ 4.582\pm 0.033$ &$\phantom{-} 0.008\pm 0.002$ &SMARTS \\ 
$3067.760$ &$  3.05$ &$ 2.480\pm 0.008$ &$ 2.350\pm 0.008$ &$ 3.256\pm 0.012$ &$ 4.573\pm 0.028$ &$\phantom{-} 0.007\pm 0.002$ &SMARTS \\ 
$3070.686$ &$  0.99$ &$ 2.460\pm 0.011$ &$ 2.337\pm 0.010$ &$ 3.232\pm 0.018$ &$ 4.538\pm 0.045$ &$\phantom{-} 0.002\pm 0.002$ &SMARTS \\ 
$3074.674$ &$  1.25$ &$ 2.471\pm 0.010$ &$ 2.311\pm 0.010$ &$ 3.241\pm 0.017$ &$ 4.569\pm 0.042$ &$\phantom{-} 0.003\pm 0.002$ &SMARTS \\ 
$3079.825$ &$ 17.12$ &$ 2.431\pm 0.005$ &$ 2.237\pm 0.005$ &$ 3.225\pm 0.007$ &$ 4.517\pm 0.015$ &$\phantom{-} 0.018\pm 0.002$ &EIGHTK \\ 
$3081.715$ &$  3.84$ &$ 2.467\pm 0.007$ &$ 2.263\pm 0.007$ &$ 3.230\pm 0.010$ &$ 4.642\pm 0.023$ &$\phantom{-} 0.008\pm 0.002$ &SMARTS \\ 
$3084.852$ &$ 13.07$ &$ 2.294\pm 0.004$ &$ 2.158\pm 0.004$ &$ 3.168\pm 0.005$ &$ 4.633\pm 0.012$ &$-0.239\pm 0.002$ & WTTM \\ 
$3088.616$ &$  3.85$ &$ 2.437\pm 0.007$ &$ 2.254\pm 0.007$ &$ 3.224\pm 0.011$ &$ 4.610\pm 0.024$ &$\phantom{-} 0.008\pm 0.002$ &SMARTS \\ 
$3092.574$ &$  3.52$ &$ 2.405\pm 0.007$ &$ 2.275\pm 0.007$ &$ 3.192\pm 0.010$ &$ 4.629\pm 0.025$ &$\phantom{-} 0.007\pm 0.002$ &SMARTS \\ 
$3103.735$ &$  2.04$ &$ 2.393\pm 0.009$ &$ 2.222\pm 0.008$ &$ 3.173\pm 0.014$ &$ 4.721\pm 0.047$ &$\phantom{-} 0.004\pm 0.002$ &SMARTS \\ 
$3104.662$ &$  6.17$ &$ 2.333\pm 0.004$ &$ 2.168\pm 0.004$ &$ 3.177\pm 0.005$ &$ 4.837\pm 0.014$ &$-0.164\pm 0.002$ & WTTM \\ 
$3107.612$ &$  4.61$ &$ 2.378\pm 0.007$ &$ 2.241\pm 0.007$ &$ 3.142\pm 0.010$ &$ 4.669\pm 0.025$ &$\phantom{-} 0.007\pm 0.002$ &SMARTS \\ 
$3111.596$ &$  2.37$ &$ 2.383\pm 0.008$ &$ 2.244\pm 0.008$ &$ 3.171\pm 0.012$ &$ 4.669\pm 0.032$ &$\phantom{-} 0.004\pm 0.002$ &SMARTS \\ 
$3114.620$ &$  3.75$ &$ 2.375\pm 0.007$ &$ 2.211\pm 0.007$ &$ 3.157\pm 0.010$ &$ 4.676\pm 0.025$ &$\phantom{-} 0.007\pm 0.002$ &SMARTS \\ 
$3129.597$ &$  1.85$ &$ 2.300\pm 0.009$ &$ 2.160\pm 0.008$ &$ 3.117\pm 0.014$ &$ 4.601\pm 0.044$ &$\phantom{-} 0.004\pm 0.002$ &SMARTS \\ 
$3132.609$ &$  0.77$ &$ 2.315\pm 0.011$ &$ 2.100\pm 0.011$ &$ 3.105\pm 0.017$ &$ 4.632\pm 0.044$ &$\phantom{-} 0.003\pm 0.002$ &SMARTS \\ 
$3133.718$ &$  5.61$ &$ 2.329\pm 0.011$ &$ 2.103\pm 0.011$ &$ 3.065\pm 0.015$ &$ 4.613\pm 0.032$ &$\phantom{-} 0.015\pm 0.002$ &EIGHTK \\ 
$3136.473$ &$  1.36$ &$ 2.321\pm 0.010$ &$ 2.086\pm 0.009$ &$ 3.074\pm 0.014$ &$ 4.646\pm 0.036$ &$\phantom{-} 0.005\pm 0.002$ &SMARTS \\ 
$3139.503$ &$  2.64$ &$ 2.266\pm 0.007$ &$ 2.099\pm 0.007$ &$ 3.060\pm 0.010$ &$ 4.667\pm 0.027$ &$\phantom{-} 0.007\pm 0.002$ &SMARTS \\ 
$3150.512$ &$  1.97$ &$ 2.254\pm 0.008$ &$ 2.087\pm 0.008$ &$ 3.079\pm 0.012$ &$ 4.692\pm 0.030$ &$\phantom{-} 0.007\pm 0.002$ &SMARTS \\ 
$3153.488$ &$  1.37$ &$ 2.222\pm 0.008$ &$ 2.110\pm 0.008$ &$ 3.115\pm 0.014$ &$ 4.708\pm 0.046$ &$\phantom{-} 0.004\pm 0.002$ &SMARTS \\ 
$3161.483$ &$  0.83$ &$ 2.248\pm 0.014$ &$ 2.089\pm 0.013$ &$ 3.059\pm 0.022$ &$ 4.601\pm 0.053$ &$-0.002\pm 0.002$ &SMARTS \\ 
$3164.466$ &$  3.56$ &$ 2.231\pm 0.007$ &$ 2.170\pm 0.007$ &$ 3.064\pm 0.011$ &$ 4.632\pm 0.027$ &$\phantom{-} 0.008\pm 0.002$ &SMARTS \\ 
$3169.555$ &$  2.90$ &$ 2.256\pm 0.008$ &$ 2.139\pm 0.008$ &$ 3.124\pm 0.013$ &$ 4.609\pm 0.029$ &$\phantom{-} 0.008\pm 0.002$ &SMARTS \\ 
$3172.505$ &$  6.09$ &$ 2.200\pm 0.007$ &$ 2.196\pm 0.007$ &$ 3.089\pm 0.010$ &$ 4.602\pm 0.027$ &$\phantom{-} 0.006\pm 0.002$ &SMARTS \\ 
$3178.485$ &$  4.37$ &$ 2.258\pm 0.011$ &$ 2.171\pm 0.012$ &$ 3.090\pm 0.018$ &$ 4.790\pm 0.049$ &$\phantom{-} 0.005\pm 0.002$ &SMARTS \\ 
$3183.473$ &$  1.43$ &$ 2.220\pm 0.009$ &$ 2.234\pm 0.009$ &$ 3.195\pm 0.016$ &$ 4.557\pm 0.038$ &$\phantom{-} 0.006\pm 0.002$ &SMARTS \\ 
$3185.513$ &$  1.16$ &$ 2.224\pm 0.009$ &$ 2.237\pm 0.010$ &$ 3.198\pm 0.017$ &$ 4.619\pm 0.043$ &$\phantom{-} 0.004\pm 0.002$ &SMARTS \\ 
$3189.471$ &$  1.36$ &$ 2.250\pm 0.012$ &$ 2.251\pm 0.013$ &$ 3.193\pm 0.022$ &$ 4.591\pm 0.050$ &$\phantom{-} 0.002\pm 0.002$ &SMARTS \\ 
$3192.472$ &$  2.30$ &$ 2.275\pm 0.010$ &$ 2.233\pm 0.011$ &$ 3.208\pm 0.017$ &$ 4.637\pm 0.037$ &$\phantom{-} 0.007\pm 0.002$ &SMARTS \\ 
$3201.474$ &$  1.26$ &$ 2.281\pm 0.010$ &$ 2.261\pm 0.011$ &$ 3.166\pm 0.017$ &$ 4.540\pm 0.036$ &$\phantom{-} 0.006\pm 0.002$ &SMARTS \\ 
$3210.475$ &$  0.53$ &$ 2.337\pm 0.032$ &$ 2.344\pm 0.036$ &$ 3.279\pm 0.071$ &$ 4.870\pm 0.231$ &$-0.030\pm 0.003$ &SMARTS \\ 
$3344.778$ &$  1.47$ &$ 2.595\pm 0.012$ &$ 2.805\pm 0.016$ &$ 3.518\pm 0.026$ &$ 4.829\pm 0.056$ &$\phantom{-} 0.002\pm 0.002$ &SMARTS \\ 
$3348.807$ &$  2.24$ &$ 2.614\pm 0.008$ &$ 2.853\pm 0.010$ &$ 3.512\pm 0.015$ &$ 4.808\pm 0.033$ &$\phantom{-} 0.006\pm 0.002$ &SMARTS \\ 
$3352.809$ &$  2.04$ &$ 2.644\pm 0.010$ &$ 2.857\pm 0.012$ &$ 3.550\pm 0.018$ &$ 4.776\pm 0.036$ &$\phantom{-} 0.006\pm 0.002$ &SMARTS \\ 
$3355.807$ &$  2.59$ &$ 2.663\pm 0.010$ &$ 2.903\pm 0.012$ &$ 3.598\pm 0.018$ &$ 4.791\pm 0.036$ &$\phantom{-} 0.006\pm 0.002$ &SMARTS \\ 
$3357.777$ &$  3.82$ &$ 2.679\pm 0.010$ &$ 2.933\pm 0.012$ &$ 3.568\pm 0.018$ &$ 4.769\pm 0.035$ &$\phantom{-} 0.007\pm 0.002$ &SMARTS \\ 
$3359.774$ &$  2.17$ &$ 2.730\pm 0.011$ &$ 2.900\pm 0.013$ &$ 3.580\pm 0.020$ &$ 4.750\pm 0.037$ &$\phantom{-} 0.006\pm 0.002$ &SMARTS \\ 
$3362.784$ &$  1.27$ &$ 2.710\pm 0.010$ &$ 2.923\pm 0.013$ &$ 3.649\pm 0.020$ &$ 4.789\pm 0.042$ &$\phantom{-} 0.003\pm 0.002$ &SMARTS \\ 
$3368.796$ &$  1.78$ &$ 2.748\pm 0.012$ &$ 2.920\pm 0.015$ &$ 3.582\pm 0.023$ &$ 4.756\pm 0.052$ &$\phantom{-} 0.004\pm 0.002$ &SMARTS \\ 
$3373.731$ &$  0.89$ &$ 2.718\pm 0.014$ &$ 2.881\pm 0.018$ &$ 3.599\pm 0.030$ &$ 4.908\pm 0.068$ &$\phantom{-} 0.003\pm 0.002$ &SMARTS \\ 
$3375.728$ &$  3.54$ &$ 2.795\pm 0.012$ &$ 2.829\pm 0.013$ &$ 3.515\pm 0.020$ &$ 4.695\pm 0.039$ &$\phantom{-} 0.005\pm 0.002$ &SMARTS \\ 
$3380.750$ &$  1.11$ &$ 2.710\pm 0.011$ &$ 2.841\pm 0.013$ &$ 3.588\pm 0.021$ &$ 4.753\pm 0.043$ &$\phantom{-} 0.002\pm 0.002$ &SMARTS \\ 
$3381.762$ &$  1.68$ &$ 2.721\pm 0.010$ &$ 2.808\pm 0.011$ &$ 3.567\pm 0.017$ &$ 4.743\pm 0.033$ &$\phantom{-} 0.005\pm 0.002$ &SMARTS \\ 
$3383.722$ &$  2.19$ &$ 2.698\pm 0.009$ &$ 2.834\pm 0.010$ &$ 3.565\pm 0.015$ &$ 4.763\pm 0.032$ &$\phantom{-} 0.004\pm 0.002$ &SMARTS \\ 
$3385.920$ &$  1.60$ &$ 2.741\pm 0.007$ &$ 2.825\pm 0.008$ &$ 3.633\pm 0.012$ &$ 5.002\pm 0.030$ &$\phantom{-} 0.007\pm 0.002$ &  APO \\ 
$3395.708$ &$  0.92$ &$ 2.703\pm 0.014$ &$ 2.842\pm 0.016$ &$ 3.649\pm 0.030$ &$ 4.828\pm 0.073$ &$-0.001\pm 0.002$ &SMARTS \\ 
$3398.745$ &$  1.20$ &$ 2.683\pm 0.012$ &$ 2.902\pm 0.014$ &$ 3.580\pm 0.023$ &$ 4.698\pm 0.051$ &$\phantom{-} 0.003\pm 0.002$ &SMARTS \\ 
$3402.736$ &$  1.98$ &$ 2.680\pm 0.009$ &$ 2.886\pm 0.010$ &$ 3.636\pm 0.016$ &$ 4.731\pm 0.035$ &$\phantom{-} 0.006\pm 0.002$ &SMARTS \\ 
$3404.833$ &$  2.90$ &$ 2.720\pm 0.010$ &$ 2.884\pm 0.012$ &$ 3.575\pm 0.017$ &$ 4.768\pm 0.035$ &$\phantom{-} 0.006\pm 0.002$ &SMARTS \\ 
$3407.773$ &$  1.46$ &$ 2.733\pm 0.011$ &$ 2.886\pm 0.013$ &$ 3.664\pm 0.020$ &$ 4.741\pm 0.035$ &$\phantom{-} 0.005\pm 0.002$ &SMARTS \\ 
$3409.800$ &$  2.36$ &$ 2.724\pm 0.009$ &$ 2.935\pm 0.010$ &$ 3.653\pm 0.015$ &$ 4.760\pm 0.029$ &$\phantom{-} 0.005\pm 0.002$ &SMARTS \\ 
$3411.826$ &$  2.11$ &$ 2.741\pm 0.010$ &$ 2.925\pm 0.012$ &$ 3.637\pm 0.018$ &$ 4.724\pm 0.033$ &$\phantom{-} 0.006\pm 0.002$ &SMARTS \\ 
$3414.809$ &$  3.83$ &$ 2.717\pm 0.009$ &$ 2.967\pm 0.011$ &$ 3.624\pm 0.016$ &$ 4.786\pm 0.033$ &$\phantom{-} 0.006\pm 0.002$ &SMARTS \\ 
$3420.807$ &$  1.28$ &$ 2.813\pm 0.012$ &$ 2.926\pm 0.014$ &$ 3.626\pm 0.021$ &$ 4.712\pm 0.036$ &$\phantom{-} 0.005\pm 0.002$ &SMARTS \\ 
$3422.774$ &$  2.40$ &$ 2.750\pm 0.010$ &$ 3.021\pm 0.013$ &$ 3.730\pm 0.020$ &$ 4.844\pm 0.044$ &$\phantom{-} 0.003\pm 0.002$ &SMARTS \\ 
$3430.752$ &$  1.98$ &$ 2.792\pm 0.012$ &$ 3.025\pm 0.016$ &$ 3.631\pm 0.023$ &$ 4.845\pm 0.051$ &$\phantom{-} 0.004\pm 0.002$ &SMARTS \\ 
$3432.882$ &$  1.51$ &$ 2.794\pm 0.015$ &$ 2.975\pm 0.020$ &$ 3.751\pm 0.032$ &$ 4.823\pm 0.059$ &$-0.001\pm 0.002$ &SMARTS \\ 
$3442.747$ &$  4.22$ &$ 2.831\pm 0.010$ &$ 3.058\pm 0.012$ &$ 3.696\pm 0.017$ &$ 4.815\pm 0.034$ &$\phantom{-} 0.007\pm 0.002$ &SMARTS \\ 
$3444.725$ &$  3.22$ &$ 2.841\pm 0.009$ &$ 3.066\pm 0.011$ &$ 3.751\pm 0.016$ &$ 4.920\pm 0.034$ &$\phantom{-} 0.006\pm 0.002$ &SMARTS \\ 
$3446.675$ &$  2.62$ &$ 2.868\pm 0.010$ &$ 3.061\pm 0.012$ &$ 3.723\pm 0.017$ &$ 4.876\pm 0.035$ &$\phantom{-} 0.006\pm 0.002$ &SMARTS \\ 
$3448.712$ &$  2.75$ &$ 2.853\pm 0.009$ &$ 3.087\pm 0.011$ &$ 3.753\pm 0.016$ &$ 4.875\pm 0.034$ &$\phantom{-} 0.004\pm 0.002$ &SMARTS \\ 
$3459.649$ &$  2.73$ &$ 2.873\pm 0.012$ &$ 3.063\pm 0.014$ &$ 3.681\pm 0.021$ &$ 5.127\pm 0.060$ &$\phantom{-} 0.006\pm 0.002$ &SMARTS \\ 
$3462.598$ &$  1.35$ &$ 2.888\pm 0.012$ &$ 3.049\pm 0.015$ &$ 3.724\pm 0.022$ &$ 4.940\pm 0.043$ &$\phantom{-} 0.005\pm 0.002$ &SMARTS \\ 
$3462.817$ &$  7.39$ &$ 2.884\pm 0.009$ &$ 3.003\pm 0.010$ &$ 3.927\pm 0.017$ &$ 5.166\pm 0.037$ &$\phantom{-} 0.002\pm 0.002$ &  APO \\ 
$3464.604$ &$  1.17$ &$ 2.921\pm 0.014$ &$ 3.024\pm 0.016$ &$ 3.729\pm 0.024$ &$ 4.955\pm 0.047$ &$\phantom{-} 0.005\pm 0.002$ &SMARTS \\ 
$3469.535$ &$  2.34$ &$ 2.925\pm 0.010$ &$ 3.083\pm 0.012$ &$ 3.740\pm 0.018$ &$ 4.975\pm 0.039$ &$\phantom{-} 0.006\pm 0.002$ &SMARTS \\ 
$3472.607$ &$  1.29$ &$ 2.970\pm 0.015$ &$ 2.986\pm 0.017$ &$ 3.645\pm 0.024$ &$ 4.894\pm 0.050$ &$\phantom{-} 0.006\pm 0.002$ &SMARTS \\ 
$3474.744$ &$  5.50$ &$ 2.904\pm 0.012$ &$ 3.000\pm 0.014$ &$ 3.714\pm 0.021$ &$ 4.991\pm 0.047$ &$\phantom{-} 0.006\pm 0.002$ &SMARTS \\ 
$3476.683$ &$  2.32$ &$ 2.932\pm 0.011$ &$ 2.968\pm 0.012$ &$ 3.699\pm 0.018$ &$ 4.902\pm 0.038$ &$\phantom{-} 0.004\pm 0.002$ &SMARTS \\ 
$3478.591$ &$  2.55$ &$ 2.879\pm 0.010$ &$ 3.011\pm 0.012$ &$ 3.715\pm 0.018$ &$ 4.893\pm 0.041$ &$\phantom{-} 0.004\pm 0.002$ &SMARTS \\ 
$3486.577$ &$  0.68$ &$ 2.867\pm 0.022$ &$ 3.006\pm 0.028$ &$ 3.680\pm 0.044$ &$ 5.166\pm 0.121$ &$-0.008\pm 0.003$ &SMARTS \\ 
$3491.630$ &$  2.38$ &$ 2.865\pm 0.010$ &$ 2.957\pm 0.012$ &$ 3.714\pm 0.018$ &$ 4.927\pm 0.039$ &$\phantom{-} 0.004\pm 0.002$ &SMARTS \\ 
$3501.538$ &$  2.77$ &$ 2.837\pm 0.012$ &$ 2.977\pm 0.014$ &$ 3.652\pm 0.020$ &$ 5.010\pm 0.051$ &$\phantom{-} 0.003\pm 0.002$ &SMARTS \\ 
$3503.538$ &$  1.20$ &$ 2.818\pm 0.013$ &$ 3.014\pm 0.015$ &$ 3.683\pm 0.023$ &$ 4.977\pm 0.057$ &$-0.001\pm 0.002$ &SMARTS \\ 
$3506.548$ &$  1.81$ &$ 2.852\pm 0.013$ &$ 2.983\pm 0.016$ &$ 3.723\pm 0.026$ &$ 5.118\pm 0.065$ &$\phantom{-} 0.005\pm 0.002$ &SMARTS \\ 
$3508.521$ &$  0.97$ &$ 2.841\pm 0.017$ &$ 3.001\pm 0.021$ &$ 3.714\pm 0.035$ &$ 4.922\pm 0.080$ &$-0.001\pm 0.002$ &SMARTS \\ 
$3511.546$ &$  1.13$ &$ 2.820\pm 0.013$ &$ 3.059\pm 0.016$ &$ 3.820\pm 0.028$ &$ 5.075\pm 0.070$ &$\phantom{-} 0.001\pm 0.002$ &SMARTS \\ 
$3512.519$ &$  1.13$ &$ 2.824\pm 0.013$ &$ 3.033\pm 0.015$ &$ 3.759\pm 0.026$ &$ 5.165\pm 0.078$ &$\phantom{-} 0.001\pm 0.002$ &SMARTS \\ 
$3520.601$ &$  1.18$ &$ 2.930\pm 0.015$ &$ 2.997\pm 0.018$ &$ 3.725\pm 0.027$ &$ 5.087\pm 0.061$ &$\phantom{-} 0.004\pm 0.002$ &SMARTS \\ 
$3521.573$ &$  4.54$ &$ 2.861\pm 0.012$ &$ 3.105\pm 0.014$ &$ 3.722\pm 0.020$ &$ 5.046\pm 0.054$ &$-0.000\pm 0.002$ &SMARTS \\ 
$3523.545$ &$  1.70$ &$ 2.929\pm 0.011$ &$ 3.036\pm 0.013$ &$ 3.758\pm 0.020$ &$ 5.017\pm 0.043$ &$\phantom{-} 0.004\pm 0.002$ &SMARTS \\ 
$3525.568$ &$  2.52$ &$ 2.913\pm 0.012$ &$ 2.993\pm 0.014$ &$ 3.712\pm 0.021$ &$ 4.996\pm 0.046$ &$\phantom{-} 0.006\pm 0.002$ &SMARTS \\ 
$3528.508$ &$  4.02$ &$ 2.893\pm 0.010$ &$ 3.050\pm 0.011$ &$ 3.752\pm 0.016$ &$ 5.079\pm 0.042$ &$\phantom{-} 0.005\pm 0.002$ &SMARTS \\ 
$3529.511$ &$  2.55$ &$ 2.911\pm 0.011$ &$ 3.012\pm 0.012$ &$ 3.729\pm 0.018$ &$ 5.087\pm 0.045$ &$\phantom{-} 0.005\pm 0.002$ &SMARTS \\ 
$3542.517$ &$  1.00$ &$ 2.855\pm 0.024$ &$ 2.837\pm 0.026$ &$ 3.653\pm 0.047$ &$ 5.104\pm 0.137$ &$-0.012\pm 0.003$ &SMARTS \\ 
$3550.558$ &$  0.60$ &$ 2.810\pm 0.027$ &$ 2.672\pm 0.026$ &$ 3.575\pm 0.048$ &$ 5.202\pm 0.155$ &$-0.019\pm 0.003$ &SMARTS \\ 
$3569.458$ &$  1.25$ &$ 2.599\pm 0.013$ &$ 2.602\pm 0.014$ &$ 3.468\pm 0.025$ &$ 5.268\pm 0.107$ &$\phantom{-} 0.001\pm 0.002$ &SMARTS \\ 
$3570.493$ &$  1.07$ &$ 2.623\pm 0.022$ &$ 2.531\pm 0.022$ &$ 3.464\pm 0.042$ &$ 5.361\pm 0.171$ &$-0.012\pm 0.003$ &SMARTS \\ 
$3578.475$ &$  1.58$ &$ 2.590\pm 0.011$ &$ 2.521\pm 0.012$ &$ 3.431\pm 0.019$ &$ 5.040\pm 0.055$ &$\phantom{-} 0.004\pm 0.002$ &SMARTS \\ 
$3580.478$ &$  1.28$ &$ 2.572\pm 0.019$ &$ 2.527\pm 0.020$ &$ 3.395\pm 0.036$ &$ 5.854\pm 0.238$ &$-0.013\pm 0.003$ &SMARTS \\ 
$3582.478$ &$  1.87$ &$ 2.607\pm 0.026$ &$ 2.458\pm 0.025$ &$ 3.352\pm 0.040$ &$ 5.370\pm 0.140$ &$-0.004\pm 0.002$ &SMARTS \\ 
$3583.461$ &$  1.12$ &$ 2.558\pm 0.024$ &$ 2.500\pm 0.026$ &$ 3.321\pm 0.044$ &$ 5.549\pm 0.220$ &$-0.014\pm 0.003$ &SMARTS \\ 

\end{longtable}
{\small {\sc Note.}  ---  HJD is the Heliocentric JD minus 2450000.  The A-D columns give the image magnitudes relative to a local reference star.  The ref. stars column gives the average magnitude differences relative to the campaign mean.} \\


\begin{deluxetable}{cccccc}
\tablecaption{{\it HST} Astrometry and Photometry of \rxj}
\tablewidth{0pt}
\tablehead{ Comp. & $\Delta\hbox{RA}$ & $\Delta\hbox{Dec}$ & $V$ & $I$ & $H$ }
\startdata
A & $\equiv 0$        & $\equiv 0$        & $18.20\pm 0.14$ & $17.77\pm0.10$ & $15.97\pm0.02$ \\
B & $+0.031\pm 0.003$ & $+1.187\pm 0.003$ & $18.25\pm 0.03$ & $18.11\pm0.11$ & $16.22\pm0.04$ \\
C & $-0.588\pm 0.001$ & $-1.120\pm 0.000$ & $18.51\pm 0.05$ & $18.19\pm0.02$ & $17.02\pm0.02$ \\
D & $-3.105\pm 0.003$ & $+0.879\pm 0.002$ & $20.02\pm 0.12$ & $19.81\pm0.03$ & $18.76\pm0.03$ \\
E & $-1.933\pm 0.003$ & $+1.139\pm 0.004$ & $24.47\pm 0.22$ & $22.73\pm0.02$ & $20.88\pm0.08$ \\
G$_{\rm d}$ & $-2.032\pm 0.002$ & $+0.586\pm 0.001$ & $19.63\pm 0.06$ & $17.88\pm0.03$ & $16.13\pm0.24$ \\
G$_{\rm e}$ & $\equiv -2.032$   & $\equiv +0.586$   & $21.72\pm 0.53$ & $19.39\pm0.35$ & $17.29\pm0.14$ \\
\enddata
\tablecomments{
Relative astrometry and absolute photometry of \rxj\ components from {\it HST}/ACS ($V$- and $I$-band) and {\it HST}/NICMOS ($H$-band) observations.  Relative positions are from the $H$-band data.   Components G$_{\rm{d}}$ and G$_{\rm{e}}$ denote the de Vaucouleurs and Exponential profiles for the galaxy model.
}
\label{tab:astromphot}
\end{deluxetable}

\begin{deluxetable}{lcccccccccccccccc}
\tablewidth{0pt}
\tablecaption{ Model Results for \rxj \label{tab:models}}
\tabletypesize{\scriptsize}
\tablehead{ Model & N$_{\rm dof}$ & $\chi^2$ & $\chi^2_{\rm pos}$ & $\chi^2_{gal}$ & $\chi^2_{del}$ &  $b$ & $G_x$ & $G_y$ & $e$ & $\theta_e$ & $\gamma$ & $\theta_{\gamma}$ & s & $\tau_{A-B}$ & $\tau_{A-C}$ & $\tau_{A-D}$ \\ & & & & & & (\arcsec) & (\arcsec) & (\arcsec) & & & & & (\arcsec) & (days) & (days) & (days) }
\startdata
  SISx        & 6 & 538  &  475  &  ---  & 63.4 & 1.86 &  $\equiv$2.032 & $\equiv$0.586 & ---         &  ---  &  0.155 & -73.6  &    ---       &   0.86 &  0.99 & -119 \\
  SIEx        & 4 & 138  &  77.2 &  ---  & 60.8 & 1.82 &  $\equiv$2.032 & $\equiv$0.586 & 0.182       & -57.0 &  0.112 & -84.9  &    ---       &   0.98 &  1.25 & -117 \\
  SIEx+       & 4 & 74.1 &   2.9 & 10.1  & 61.0 & 1.83 &          2.036 &         0.571 & 0.162       & -59.1 &  0.113 & -82.6  &    ---       &   0.98 &  1.23 & -117 \\
              &   &      &       &       &      &      &                &               &             &       &        &        &              &        &       &      \\
  SIEx+       & 2 & 23.8 &  0.1  & 0.2   & 23.5 & 1.55 &          2.033 &         0.585 & 0.132       & -62.6 &  0.035 &  81.7  &    ---       &   5.90 &  7.88 &  -80 \\
$\alpha=1$    &   &  --- &  ---  & ---   & ---  & 0.32 &         -0.049 &        -0.035 & $\equiv 0$  &  ---  &   ---  &  ---   &   0.16       &   ---  &  ---  &  --- \\
              &   &      &       &       &      &      &                &               &             &       &        &        &              &        &       &      \\
 SIEx+        & 0 &  4.7 &  0.2  & 0.2   &  4.3 & 1.66 &          2.032 &         0.586 & 0.165       & -63.4 &  0.052 &  87.4  &    ---       &  12.00 & 13.78 &  -85 \\
$\alpha=0.72$ &   &  --- &  ---  & ---   & ---  & 0.27 &         -0.056 &        -0.041 & $\equiv 0$  &  ---  &   ---  &  ---   & $\equiv 0.2$ &   ---  &  ---  &  --- \\

\enddata
\tablecomments{Best-fit model parameters for five models described in the text.  For the $\alpha=1$ and $\alpha=0.72$ perturber models, the first line gives the model parameters for the primary SIEx+ halo and the second line gives the model parameters for the perturber near image A.  All relative positions are measured with respect to image A.}
\end{deluxetable}

\end{document}